\documentclass[twocolumn]{aastex63}

\shorttitle{Li Evolution in NGC 2243}
\shortauthors{Anthony-Twarog, Deliyannis, Twarog}

\begin{document}


\title{WIYN Open Cluster Study LXXXV. Li in NGC 2243 - \\
Implications for Stellar and Galactic Evolution}

\author{Barbara J. Anthony-Twarog}
\affiliation{Department of Physics and Astronomy, University of Kansas, Lawrence, KS 66045-7582, USA}
\email{bjat@ku.edu}

\author{Constantine P. Deliyannis}
\affiliation{Department of Astronomy, Indiana University, Bloomington, IN 47405-7105, USA}
\email{cdeliyan@indiana.edu}

\author{Bruce A. Twarog}
\affiliation{Department of Physics and Astronomy, University of Kansas, Lawrence, KS 66045-7582, USA}
\email{btwarog@ku.edu}



\begin{abstract}
High-dispersion spectra in the Li 6708 \AA\  region have been obtained and analyzed in the old, metal-deficient cluster, NGC 2243. From Hydra spectra for 29 astrometric and radial-velocity members, we derive rotational velocities, as well as [Fe/H], [Ca/H], [Si/H], and [Ni/H] based on 17, 1, 1, and 3 lines, respectively.  Using ROBOSPECT, an automatic equivalent width measurement program, we derive [Fe/H] $= -0.54 \pm$ 0.11 (MAD), for an internal precision for the cluster [Fe/H] below 0.03 dex.  Given the more restricted line set, comparable values for [Ca/H], [Si/H], and [Ni/H] are $-0.48 \pm$ 0.19, $-0.44 \pm$ 0.11, and $-0.61 \pm$ 0.06, respectively. With E$(B-V)$ = 0.055, appropriate isochrones imply $(m-M)$ = 13.2 $\pm$ 0.1 and an age of 3.6 $\pm$ 0.2 Gyr.  Using available VLT spectra and published Li abundances, we construct a Li sample of over 100 stars extending from the tip of the giant branch to 0.5 mag below the Li-dip. The Li-dip is well populated and, when combined with results for NGC 6819 and Hyades/Praesepe, implies a mass/metallicity slope of 0.4 M$_{\sun}$/dex for the high mass edge of the Li-dip. The A(Li) distribution among giants reflects the degree of Li variation among the turnoff stars above the Li-dip, itself a function of stellar mass and metallicity and strongly anticorrelated with a $v_{rot}$ distribution that dramatically narrows with age. Potential implications of these patterns for the interpretation of Li among dwarf and giant field populations, especially selection biases tied to age and metallicity, are discussed.
\end{abstract}


\section{Introduction}

With improved precision and rapidly expanding databases due to the application of larger telescopes in survey mode (see, {\it e.g.} \citet{CA19, GA19}), the abundance of atmospheric Li (A(Li)\footnote{A(Li)=log$N_{Li}$ - log$N_{H}$ + 12.00}) has become an increasingly valuable signature of both stellar structure and galactic chemical evolution. Since lower mass main sequence and red giant stars represent a crucial interface between both research
areas, an extensive spectroscopic program has been underway to survey members of a key set of open clusters from the tip of the giant branch to as far down the main sequence as the technology allows. New spectroscopy 
and/or reanalyses of published data have been discussed for NGC 752, NGC 3680, and IC 4651 \citep{AT09}, NGC 6253 \citep{AT10, CU12}, NGC 2506 \citep{AT16, AT18}, Hyades and Praesepe \citep{CU17}, NGC 6819 \citep{AT14, LB15, DE19}, and, most recently, NGC 188 \citep{SU20}. The regularly repeated rationale for investigation of these clusters and their role within the bigger picture of Li evolution can be found in the respective papers and will not be detailed here. An overview of the goals and some current insight into the benefits of a more comprehensive approach to open cluster analyses can be found in \citet{TW20}. 

The focus of this investigation is the metal-deficient, older open cluster, NGC 2243. Just how metal-deficient the cluster is will be a point we return to in Section 3. There is, however, longstanding agreement that, irrespective of
the technique used to derive the metallicity and/or the exact zero point of the metallicity scale, when ranked by metallicity among the well-studied open clusters within 5 kpc of the Sun, NGC 2243 sits consistently at or near the bottom of the list \citep{TW97, FR02, NE16}. This combination of distance, metallicity, and an age competitive with that of M67 (see Section 4) makes NGC 2243 an important testbed for the effects of metallicity on the rate and degree of Li evolution among lower mass stars while on the main sequence  
and, to an even greater extent, in subgiant and giant branch phases. 
Isolating the metallicity-dependent effects of atmospheric evolution also allows exposure of the underlying patterns of Galactic Li evolution at both the metal-poor and metal-rich ends of the scale \citep{ra20}. Moreover, for now, the main sequence stars provide one of the closest links to their analogs within the halo population and a constraint on the apparent contradiction between stellar and cosmological predictions for the primordial A(Li) \citep{SP82a,SP82b}.

In an attempt to expand the spectrosopic sample over the full range of luminosity from the tip of the red giant branch to the main sequence below the Li-dip \citep{BT86}, our own Hydra data, detailed in Section 2, have been supplemented when possible by high resolution spectra obtained at the VLT, and by reconsidered A(Li) from published sources when no spectra were accessible. The layout of the discussion is as follows: 
Section 2 details the new spectra, their processing, reduction, and velocity derivation, as well as the additional spectra from other sources, and a compilation of precision broad-band photometry for color-magnitude diagram (CMD) analysis; Section 3 lays out the cluster metallicity determination from the Hydra spectra and 
comparison with previous results; Section 4 combines the metallicity, reddening, and $BV$ CMD to define the cluster age and distance, and indirectly the masses of the stars populating the turnoff and giant branch, through comparison to appropriate isochrones; Section 5 describes the Li abundance estimation for new members within the Hydra and VLT samples and combines these with previously published determinations to delineate the evolutionary trend within the cluster and place the cluster within the context of Li changes with both age and metallicity. Section 6 summarizes our results for NGC 2243 while illuminating the impact of the varying boundaries of the Li-dip due to metallicity effects on our understanding of Galactic Li evolution and selection bias.

\section{Observational Data: New, Old, and Revisited}

\subsection{Spectroscopy: WIYN Hydra Data - Acquisition and Reduction}

NGC 2243 is a compact cluster and, while rather far south for northern hemisphere observations, access to the Hydra multi-object spectrograph on the WIYN 3.5-meter telescope\footnote{The WIYN Observatory was a joint facility of the University of Wisconsin-Madison, Indiana University, 
Yale University, and the National Optical Astronomy Observatory.} motivated spectroscopic observations in this critical cluster.

Our candidate list of cluster members was constructed long before $Gaia$ DR2 \citep{GA18} and was based primarily upon analyses of precision 
extended Str{\"o}mgren photometry \citep{AT05} (ATAT) using the traditional intermediate and narrow-band indices of the system. Under the assumption that cluster members should exhibit similar chemical composition and reddening, we additionally 
compiled a new parameter $H$, defined as $hk -2.25(b-y)$, and compared this photometric construct to then current membership information, mainly radial velocities.  $H$ values for the most likely member candidates centered on $-0.40$; any star that deviated significantly from this target figure was eliminated as a spectroscopic target. Table 1 lists basic information about our Hydra sample, including coordinates, WEBDA\footnote{https://webda.physics.muni.cz} identification numbers, and $V$ and $B-V$ photometry compiled 
as described below.  All of the stars above the dividing horizontal line in Table 1 are judged members by \citet{CA18} based on $Gaia$ DR2 astrometry. Additional data about each star's rotational and radial velocities are presented, with further descriptions to follow. 

Two Hydra fiber configurations were constructed, one with nine bright stars and another with an eventual total of 33 fainter targets designed for longer, multiple exposures.  The brighter configuration was observed for three 30-minute exposures on 21 Jan., 2015, where the date indicates the UT date.  Stars in the fainter configuration were observed with exposures ranging from 53 to 90 minutes, for 180 minutes total on 21 Jan., 2015, 268 minutes on 22 Jan., 2015 
and 246 minutes on 12 Jan., 2016, amounting to 11.6 hours of total exposure.  Between the 2015 and 2016 runs, one fiber fell out of use so that one star, observed only in 2015, was replaced by another star observed only in 2016; the other 31 stars were observed in both years. Notes to Table 1 indicate which stars were observed in only one epoch.

\floattable
\begin{deluxetable}{rrrrrrrrrrr}
\tablenum{1}
\tablecaption{Stellar Characteristics, Hydra Sample}
\tablewidth{0pt}
\tablehead{
\colhead{WEBDA ID} & \colhead{$\alpha (2000)$} & \colhead{$\delta (2000)$} & \colhead{$V$} & \colhead{$B-V$} & \colhead{S/N} & 
\colhead{$V_{rad}$} & \colhead{$\sigma_{Vrad}$} & \colhead{$v_{rot}$} & \colhead{$\sigma_{vrot}$} & \colhead{Note} \\
\colhead{} & \colhead{} & \colhead{} & \colhead{} & \colhead{} & \colhead{per pix.} & 
\colhead{km-s$^{-1}$} & \colhead{km-s$^{-1}$} & \colhead{km-s$^{-1}$} & \colhead{km-s$^{-1}$} } 
\startdata
519 & 97.307140 & -31.350243 & 15.724 & 0.524  & 157  & 55.54 & 1.89 & 16.54 & 1.03 & \\
611 & 97.317848 & -31.279542 & 15.384 & 0.861  & 185  & 55.67 & 1.64 & 11.92 & 0.69 & \tablenotemark{a} \\
716 & 97.327288 & -31.225097 & 15.731 & 0.504  & 154  & 44.64 & 2.18 & 19.56 & 1.32 & \\
873 & 97.342210 & -31.281069 & 15.712 & 0.557  & 164  & 55.20 & 1.35 & 8.86 & 0.43 & \\
1044 & 97.357060 & -31.273251 & 15.702 & 0.493 & 135  & 53.16 & 2.13 & 13.56 & 0.93 & \\
1133 & 97.362952 & -31.307905 & 16.074 & 0.497 & 60   & 39.72 & 3.60 & 31.96 & 2.82 & \tablenotemark{a}\\
1230 & 97.369157 & -31.205316 & 15.996 & 0.744 & 153  & 54.80 & 1.57 & 11.96 & 0.67 & \\
1263 & 97.372145 & -31.295469 & 15.241 & 0.807 & 196  & 55.74 & 1.70 & 21.76 & 1.10 & \tablenotemark{a} \\
1266 & 97.372710 & -31.366009 & 15.662 & 0.549 & 124  & 54.74 & 1.84 & 14.65 & 0.92 & \tablenotemark{a} \\
1271 & 97.372511 & -31.262745 & 13.692 & 0.918 & 132  & 56.95 & 0.73 & 11.70 & 0.29 &  \tablenotemark{a,b}\\
1294 & 97.374103 & -31.273084 & 15.472 & 0.861 & 127  & 56.16 & 1.38 & 11.23 & 0.44 & \tablenotemark{a,b}\\
1313 & 97.375413 & -31.282930 & 12.887 & 1.097 & 124  & 57.51 & 0.72 & 4.65 & 0.12 & \tablenotemark{a,c}\\
1421 & 97.381966 & -31.259343 & 15.868 & 0.826 & 153  & 55.82 & 2.17 & 14.84 & 1.07 & \\
1436 & 97.382878 & -31.301186 & 15.823 & 0.480 & 134  & 53.69 & 2.18 & 19.17 & 1.32 & \\
1467 & 97.385029 & -31.291483 & 13.713 & 0.930 & 130  & 56.12 & 0.83 & 15.61 & 0.44 & \tablenotemark{a,b}\\
1696 & 97.396588 & -31.289930 & 14.201 & 0.965 & 199  & 57.63 & 1.24 & 11.96 & 0.52 & \\
1738 & 97.398569 & -31.286365 & 13.721 & 0.925 & 129  & 56.64 & 0.86 & 18.95 & 0.51 & \tablenotemark{b}\\
1847 & 97.403533 & -31.245067 & 14.219 & 0.962 & 164  & 58.18 & 1.34 & 16.23 & 0.72 & \\
1871 & 97.405196 & -31.318901 & 15.485 & 0.482 & 159  & 55.34 & 1.89 & 15.36 & 0.96 & \\
1995 & 97.412396 & -31.287251 & 15.909 & 0.824 & 127  & 54.79 & 1.47 & 10.94 & 0.55 & \\
2003 & 97.412831 & -31.299608 & 15.731 & 0.469 & 138  & 53.36 & 2.99 & 21.70 & 1.91 & \\
2098 & 97.417557 & -31.247321 & 15.884 & 0.707 & 129  & 56.00 & 1.47 & 11.14 & 0.55 & \\
2135 & 97.421203 & -31.318307 & 13.802 & 1.072 & 315  & \nodata & 1.67 & 38.77 & 1.54 & \tablenotemark{a,d}\\
2394 & 97.439450 & -31.250351 & 15.726 & 0.540 & 137  & 55.30 & 2.57 & 13.01 & 1.37 & \\
2410 & 97.440970 & -31.260588 & 13.633 & 0.936 & 105  & 58.43 & 0.90 & 12.67 & 0.39 & \tablenotemark{a,b}\\
2434 & 97.442601 & -31.267117 & 14.076 & 0.821 & 248  & 55.42 & 1.34 & 11.73 & 0.54 & \\
2676 & 97.464961 & -31.303919 & 15.996 & 0.452 & 123  & 53.02 & 2.19 & 12.76 & 1.00 & \\
2696 & 97.466663 & -31.238735 & 16.007 & 0.451 & 130  & 54.05 & 3.07 & 20.15 & 1.85 & \\
2908 & 97.488997 & -31.267521 & 16.021 & 0.455 & 113  & 56.42 & 2.84 & 20.71 & 1.80 & \\
3618 & 97.422947 & -31.243319 & 13.683 & 1.046 & 107  & 56.60 & 0.71 & 11.54 & 0.27 & \tablenotemark{a,b}\\
3633 & 97.368411 & -31.288134 & 12.025 & 1.418 & 81   & 57.11 & 1.07 & 13.13 & 0.46 & \tablenotemark{a,b}\\
3728 & 97.295890 & -31.344250 & 13.316 & 0.992 & 119  & 57.79 & 0.96 & 19.58 & 0.59 & \tablenotemark{a,b} \\
\hline
259 & 97.272604 & -31.284006 & 14.714 & 0.957 & 175 & 58.33 & 1.20 & 9.70 & 0.41 & \\
507 & 97.305208 & -31.280300 & 12.112 & 1.048 & 130 & 80.21 & 0.88 & 9.01 & 0.28 & \tablenotemark{b}\\
552 & 97.310604 & -31.234522 & 14.820 & 0.508 & 249 & 46.93 & 1.21 & 12.58 & 0.55 & \\
718 & 97.327242 & -31.188708 & 15.862  & 0.519 & 149 & 74.54 & 3.00 & 22.14 & 2.17 & \\
2451 & 97.444504 & -31.384656 & 13.985 & 0.594 & 248 & -13.58 & 1.94 & 13.93 & 1.07 & \\
2528 & 97.450467 & -31.324308 & 15.736  & 0.539 & 121  & 67.83 & 2.44 & 24.58 & 1.67 & \\
2704 & 97.467308 & -31.281325 & 15.070   & 0.967 & 203  & 3.74 & 1.84 & 16.44 & 0.98 & \\
2719 & 97.468896 & -31.359297 & 13.881 & 0.531 & 282  & 3.90 & 1.62 & 17.18 & 0.92 & \\
3063 & 97.510746 & -31.266553 & 15.284 & 0.560 & 147  & 45.06 & 1.41 & 15.04 & 0.78 & \\
3726 & 97.304417 & -31.309617 & 11.576 & 1.166 & 36  & 117.27 & 1.10 & 10.39 & 0.39 & \tablenotemark{b}\\
\enddata
\tablenotetext{a}{Li analysis from additional VLT spectra}
\tablenotetext{b}{Radial velocity obtained from only 2015 spectra.}
\tablenotetext{c}{Radial velocity obtained from only 2016 spectra.}
\tablenotetext{d}{Radial velocity for W2135 was 39.7 km-s$^{-1}$ in 2015, 74.4 km-s$^{-1}$ in 2016.}
\end{deluxetable}

Our spectrograph setup is designed to cover a wavelength range $\sim$400 \AA\ centered on 6650 \AA, with dispersion of 0.2 \AA\ per pixel. Examination of Thorium-Argon lamp spectra indicates a line resolution comprising 2.5 pixels, yielding a spectral resolution over 13,000.  The data were subsequently processed using standard reduction routines in IRAF\footnote{IRAF is distributed by the National Optical Astronomy Observatory, which is operated by the Association of Universities for Research in Astronomy, Inc., under cooperative agreement with the National Science Foundation.}, including, in order of application, bias subtraction, division by the averaged flat field, dispersion correction through interpolation of the comparison lamp spectra, throughput correction for individual fibers using daytime sky 
exposures in the same configuration, and continuum normalization. After flat field division and before the dispersion correction, the long-exposure program images were cleaned of cosmic rays using ``L. A. Cosmic''\footnote{http://www.astro.yale.edu/dokkum/lacosmic/, an IRAF 
script developed by P. van Dokkum (van Dokkum 2001); spectroscopic version.}
\citep{VD01}. Real-time sky subtraction was accomplished by using the dozens of fibers not assigned to stars and exposed to the sky during each integration.  

Cumulative spectra for each configuration were constructed by additive combination by night and by year; the differences between 2015 and 2016 values were 
also examined for signs of potential radial-velocity variation before construction of cumulative combination spectra.  Other than WEBDA 2135 (W2135), discussed in detail in \citet{AT20}(Paper I), no other stars show significant variation between the two epochs. The signal-to-noise ratio per pixel (S/N) has been estimated by direct inspection of the composite spectra within IRAF's SPLOT utility, using mean values and r.m.s. scatter from a relatively line-free region between 6680 and 6694 \AA.  S/N estimates of Table 1 characterize the summed composite spectra, with a median value of 138 achieved for the sample. 

\subsection{Spectroscopy: VLT Data}
Largely because early analyses indicated a fascinating range of anomalies for member giant W2135 (Paper I), we sought to expand our sample 
of spectroscopically observed stars by searching for archival spectra within the field of NGC 2243 and with  spectral coverage in the region of the Li line. We gratefully acknowledge use of the ESO Science Archive Facility\footnote{http://archive.eso.org/scienceportal} from which we retrieved fully processed spectra obtained for the public {\it Gaia} ESO Survey at the Very Large Telescope (VLT) with the Fibre Large Array Multi Element Spectograph (FLAMES) fiber-feed assembly to both the high resolution UVES spectrograph and the GIRAFFE spectrograph, with pixel-resolutions of 16.9 m\AA\ and 50 m\AA\, respectively. As our intention was to use these spectra for Li abundance analysis only, we performed no further processing other than to import them to a format acceptable to standard IRAF routines. The higher resolution of the VLT spectra, particularly those from the UVES instrument, provides a considerable advantage for Li analysis although S/N above 100 is still desirable even for higher resolution spectra.  Our initial archive search was limited to spectra described as having S/N above 70; where possible, multiple exposures were co-added to produce spectra with higher S/N.

\floattable
\begin{deluxetable*}{rrrrrrrrrr}
\tablenum{2}
\tablecaption{Stellar Characteristics, VLT Sample}
\tablewidth{0pt}
\tablehead{
\colhead{WEBDA ID} & \colhead{$\alpha (2000)$} & \colhead{$\delta (2000)$} & \colhead{$V$} & \colhead{$B-V$} & 
\colhead{S/N} & \colhead{T$_{eff}$} &  \colhead{log $g$} & \colhead{$\xi$} } 
\startdata
239  & 97.269875 & -31.357306 & 13.644 & 0.891 & 94  & 4990 & 2.5  & 1.50 \\
365  & 97.288917 & -31.175694 & 14.009 & 0.975 & 151 & 4824 & 2.5  & 1.50 \\ 
910  & 97.345833 & -31.291639 & 13.717 & 0.916 & 196 & 4939 & 2.5  & 1.50 \\
1261 & 97.372208 & -31.359833 & 15.177 & 0.873 & 130 & 5027 & 3.1  & 1.39 \\
1679 & 97.395958 & -31.322167 & 15.113 & 0.874 & 99  & 5025 & 3.1  & 1.39 \\
1686 & 97.396333 & -31.338639 & 15.675 & 0.496 & 130 & 6324 & 3.8  & 1.81 \\
1923 & 97.408167 & -31.276917 & 15.838 & 0.449 & 114 & 6538 & 3.9  & 1.97 \\
2195 & 97.424500 & -31.263194 & 15.321 & 0.766 & 133 & 5267 & 3.2  & 1.41 \\
2536 & 97.451542 & -31.377111 & 15.234 & 0.892 & 135 & 4987 & 3.1  & 1.38 \\
2613 & 97.458250 & -31.278306 & 15.146 & 0.828 & 94  & 5124 & 3.1  & 1.41 \\
2648 & 97.462536 & -31.245210 & 14.449 & 0.942 & 88  & 4888 & 2.75 & 1.50 \\
\enddata
\end{deluxetable*}

As a result, additional spectra for some of the stars listed in Table 1 were found and examined to take advantage of the higher resolution of the VLT spectra.  Notes to Table 1 indicate the Hydra sample stars for which VLT spectra were also examined.  VLT spectra for eleven probable member stars not included in the Hydra sample were also identified.  Table 2 presents the basic information for these stars, comparable to that presented for the Hydra stars in Table 1, although radial and rotational velocity values have not been estimated for the VLT spectra; in the place of these data, atmospheric parameters used in later analysis steps, are included in this table. S/N values are reproduced from the processed spectra's headers or report quadratic sums for co-added spectra. Direct measurements of the S/N between 6680 and 6694 \AA\ largely confirm these listed values as adequate representation of the S/N per pixel.

\subsection{WIYN Hydra Data: Radial and Rotational Velocities}
We utilized the {\it fxcor} Fourier-transform, cross-correlation utility in IRAF to assess velocites for each star in each year's composite spectra. In {\it fxcor}, stars are compared to a stellar template of similar $T_{\mathrm{eff}}$ for the wavelength range of our spectra redward of H$\alpha$. Output of the 
{\it fxcor} utility characterizes the cross correlation function (CCF), from which estimates of each star's radial velocity are easily inferred. 

From the procedure for radial-velocity estimation, rotational velocities 
can also be estimated from the cross correlation function full-width half maximum (CCF FWHM) using a procedure developed by \citet{ST03}. 
This procedure exploits the relationship between the CCF FWHM, line widths and $v_{rot}$, using a set of 
numerically ``spun up" standard spectra with comparable spectral types to constrain the relationship. Our analyses of both red giant and turnoff spectra
demonstrate that $fxcor$ cross-correlation profiles have significantly reduced accuracy when attempting to reproduce rotational velocities above 30 km-s$^{-1}$.  Our spectral resolution implies that rotational velocities below 10 km-s$^{-1}$ are not meaningful.

Radial and rotational velocities for each of our program stars are included in Table 1, divided into likely members and nonmembers based on $Gaia$ astrometry. Listed errors reflect the precision of the template fit by {\it fxcor}.  
As noted earlier, the lines of Table 1 are separated by astrometric membership categories applied {\it after} sample selection and data acquisition.  
The top 32 stars are considered members in the compilation by \citet{CA18} based on $Gaia$ DR2 astrometry, while the bottom ten stars, initially selected by photometric 
criteria to be potential members, are now designated as unlikely astrometric members. In nearly every case, our radial-velocity information for the ten likely nonmembers is consistent with the astrometric classification. 
The exception is star W259 for which DR2 astrometry indicates a parallax $\sim 1/2$ the cluster average value and a total proper motion $\sim 1/3$ the magnitude of the cluster motion \citep{CA18}.
In addition to the ten astrometric nonmembers, stars W2135, W1133, and W716 have been excluded from the computation of the average cluster radial velocity for reasons addressed below. An average radial velocity from the most likely members is $55.8 \pm 1.5$ km-s$^{-1}$, where the error statistic is the standard deviation among the 29 stars. 

Star W2135 has been discussed in detail in Paper I. In summary, in addition to a measurable (and therefore distinctive for the cluster's age and the star's evolutionary state) Li line, this giant had been flagged as a photometric variable by ATAT and as V14 by \citet{KA06}.  Limited information regarding photometric amplitude or periodicity emerged from either photometric study, beyond a period estimated in days rather than hours. From Hydra spectra, we were only able to note a significant change in radial velocity between our two observational epochs, from $39.7$ km-s$^{-1}$ in 2015 to $74.4$ km-s$^{-1}$ one year later, for an average fortuitously near the cluster mean.  Our velocity analyses indicated abnormally large $v_{rot}$ sin $i$ for a giant star, at least 40 km-s$^{-1}$, although we note again that analysis of the $fxcor$ cross-correlation profiles probably can't accurately reproduce rotational velocities above $\sim$30 km-s$^{-1}$. 

Reiterating a note added in proof to Paper I, the All-Sky Automated Survey for SuperNovae (ASAS-SN) variable stars data base \citep{SH14}\footnote{https://asas-sn.osu.edu/variables} includes star W2135.  The cataloged star, ASASSN-V J062941.11-311906.9/V0412 CMa, has a variability class of ROT, implying that the variability is likely due to rotationally-induced brightness modulation, with a period of 6.755421 days \citep{JSK}. Thus, the estimated true $v_{rot}$, {\it i.e.} with the sin $i$ effect
removed, using stellar parameter estimates derived from the CMD location of the star becomes 86 km-s$^{-1}$ (Paper I).

The spectra for W1133 are essentially featureless other than H$\alpha$, indicating either a higher temperature than suggested by the star's color, rapid rotation, or both. This star is also a variable candidate from \citet{KA06}(V13), with a tentative classification as a $\gamma$ Doradus variable and 
a period of $\sim 0.77$ days.  We were unable to obtain $fxcor$ results for one of the two epochs and consider its one radial-velocity determination tentative. Star W716 had a consistent radial velocity indicative of non-membership for both epochs, in spite of a classification as a highly probably member based on astrometry. If a cluster member, it must be a single-lined spectroscopic binary (SB1) and we were unlucky in our phase sampling, a result confirmed below.  

\subsection{Comparison With Other Radial-velocity Determinations}
A few radial-velocity standards were observed on some, but not all, of the 2015 and 2016 nights on which NGC 2243 was observed.  Our radial velocities derived from $fxcor$ analysis yield $V_{rad}$ that are, on average, $0.6 \pm 0.7$ km-s$^{-1}$  too large in comparison to the velocities quoted for our adopted standards on the SIMBAD database\footnote{operated at CDS, Strasbourg, France \citep{WE00}, http://simbad.u-strasbg.fr/simbad}. Although the absolute scale of our radial velocities was considered less critical than the internal or relative values for purposes of confirming common membership or detecting SB candidates, there are several radial-velocity surveys of NGC 2243 with which we can compare our mean results, even though the direct overlap in stellar samples is significant for only two surveys. Comparisons of the radial velocities of overlapping stars to other past surveys have been constructed and are summarized in Table 3. 

\citet{FR13}(FR13) derive a cluster mean radial velocity of 61.9 km-s$^{-1}$, some 6.1 km-s$^{-1}$  higher than our Hydra-based result of $+55.8$ km-s$^{-1}$ from the 29 member stars. 
Only one star is common to both studies, the likely nonmember W552; our derived radial velocity is 5.2 km-s$^{-1}$  smaller than that quoted by FR13. The most recent, and largest, analysis of motions in the cluster by \citet{G3D}(G3D) indicates a cluster velocity of $59.82 \pm 0.58$ km-s$^{-1}$, again larger than our average value by 4.2 km-s$^{-1}$, where the quoted error reflects only the internal dispersion within the cluster.

\floattable
\begin{deluxetable}{cccc}
\tablenum{3}
\tablecaption{Radial-Velocity Comparisons: Table 1 - LIT}
\tablewidth{0pt}
\tablehead{
\colhead{Survey} & \colhead{$\Delta(V_{rad})$} & \colhead{$\sigma$} & \colhead{Number} \\
\colhead{} & \colhead{km-s$^{-1}$} & \colhead{km-s$^{-1}$} & \colhead{of stars}  } 
\startdata
$Gaia$ DR2 & -1.7 & 1.3 & 6 \\
$Gaia$ 3D & -4.2 & 1.8 & 24 \\
FR13 & -5.2 & \nodata & 1 \\
JFP & 0.2 & 1.3 & 12 \\
Gratton & -3.7 & 1.1 & 2 \\
Mermilliod & -2.9 & \nodata & 1 \\
FJ93 & -8.2 & 10.4 & 4 \\
FR02 & -4.1 & 12.4 & 7 \\
Kaluzny & 2.8 & \nodata & 1 \\
Minitti & 2.8 & 9.6 & 7\\
Cameron \& Reid & -16.9 & 9.9 & 5\\
\enddata
\tablecomments{DR2: \citet{GA18}; G3D: \citet{G3D}; FR13: \citet{FR13}; JFP: \citet{JA11}; \citet{GR82}; \citet{Merm}; FJ93: \citet{FJ93}; FR02: \citet{FR02};
Kaluzny: \citet{K96}; \citet{Mi95}; \citet{CR87}. }
\end{deluxetable}

We have referred to the G3D analysis to re-evaluate the membership credentials of stars described in Tables 1 and 2 (Hydra and VLT samples), as well as samples reported on by FR13 and \citet{HI00}(HP). Referring once again to stars excluded from our Hydra radial-velocity analysis, radial-velocity discrepancies for stars W2135, W1133 and W716 are confirmed.  In Table 1, $V_{rad}$ for W716 is noted as 44.6 km-s$^{-1}$, quite a bit lower than the internal average for the cluster of 55.8 km-s$^{-1}$.  Results from G3D indicate a radial velocity of 68.3 km-s$^{-1}$, but likely membership, suggesting even more strongly an SB1 nature for this star. The situation is much the same for W1133, known to be a variable and potential binary; the G3D $V_{rad}$ is even lower than the 39.7 km-s$^{-1}$ listed in Table 1. Finally, W2135 exhibits a G3D radial velocity of 80.4 km-s$^{-1}$, similar to our larger 2016 measurement and further confirmation of its SB character.

\subsection{Photometry and Reddening}
Reliable photometry plays a dual role in the discussion of NGC 2243. When adjusted for metallicity and reddening, it (a) is the basis for deriving the $T_{\mathrm{eff}}$ used in spectroscopic abundance estimation and, (b) permits age and distance determination through comparison to appropriate isochrones, leading to mass estimates for individual stars at a variety of CMD positions. A comprehensive discussion of the broad-band $BV$ photometry available at the time can be found in ATAT. For NGC 2243, only two significant CCD $BV$ surveys of the cluster were then available \citep{BO90, BE91}, focusing on the central region of the cluster due to smaller format CCDs. Fortunately, comparison of the two data sets showed excellent agreement, indicating that both were well tied to the standard $BV$ system at the time as defined by \citet{LA83} and \citet{JG82}. The merger of these two photometric sets is described in detail in ATAT and will not be repeated. Somewhat surprisingly, with the exception of the $Gaia$ DR2 dataset, no wide area broad-band studies, $BV$ or otherwise, have been added to the literature since then.

\begin{figure}
\figurenum{1}
\plotone{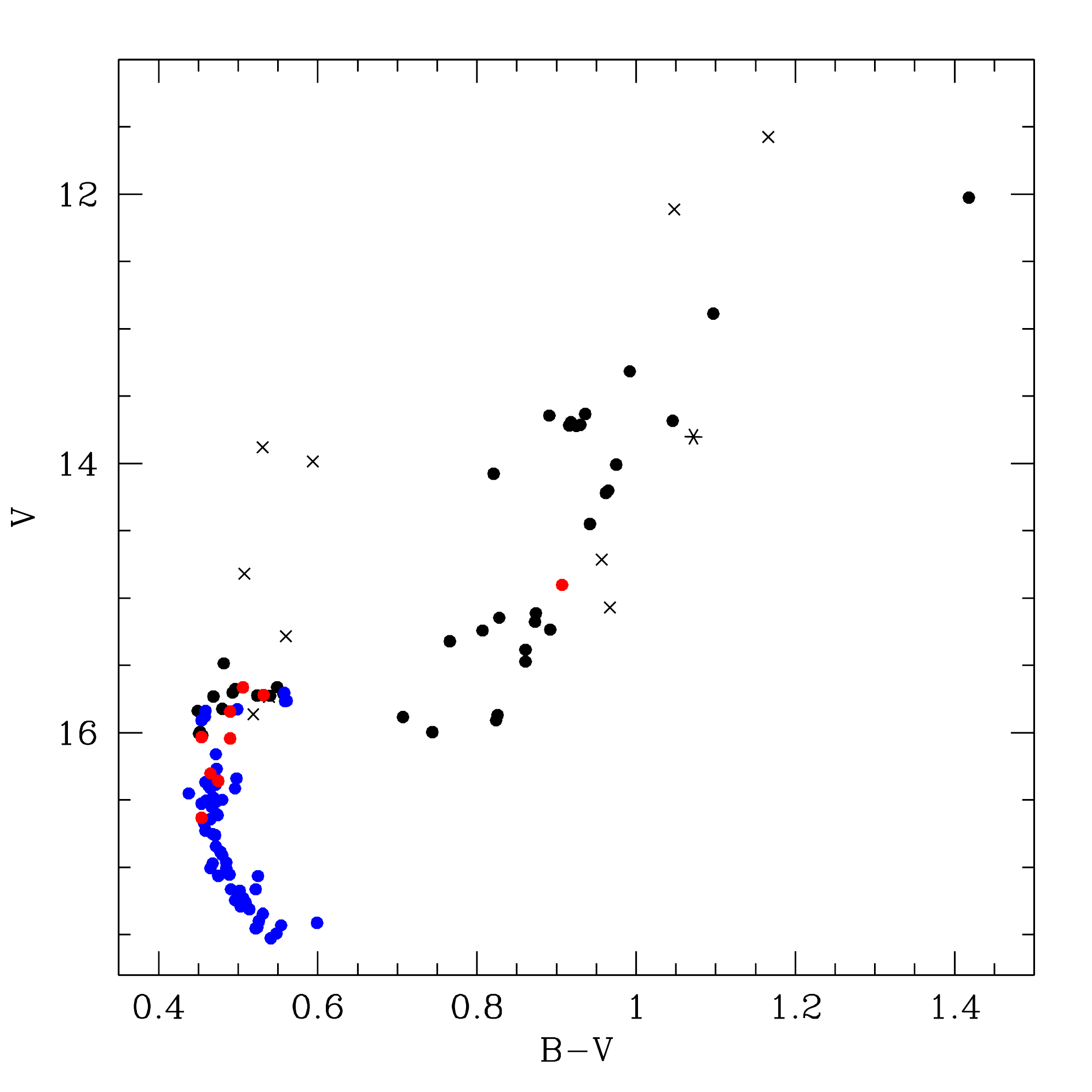}
\caption{CMD of stars with Li spectroscopy. Filled black circles and crosses are the members and nonmembers, respectively, from Tables 1 and 2. Filled blue circles are the members from FR13, and the filled red circles are members from HP. W2135 is noted with an asterisk symbol. }
\end{figure}

For our purposes the photometric compilation has been broken down into two distinct regions, the red giants ($(B-V) \geq 0.70$) and the turnoff region (($B-V) < 0.70$) through the unevolved main sequence stars to $V$ $\sim$ 18.5. For the red giants, the primary photometric data set is the precision $uvbyCa$H$\beta$ CCD photometry of ATAT. ATAT used the multicolor indices to identify highly probable cluster members and to convert the high precision ($V$, $b-y$) for likely members to the traditional ($V$, $B-V$) system defined by the earlier, smaller CCD samples \citep{BE91, BO90}. As detailed in Paper I, the $Gaia$ DR2 ($G$, $B_{p}-R_{p}$) photometry for the red giants was readily transferred to the ($V$, $B-V$) system using 25 member giants \citep{CA18} in common, with the scatter in residuals between the two systems measured at $\pm$0.007 mag and $\pm$0.009 mag in $V$ and ($B-V$), respectively. This allowed a merger of the two photometric datasets and reliable expansion of the $BV$ system for giants to $Gaia$ DR2 members outside the field studied by ATAT. As detailed in Paper I, this leads to 39 red giant members in the color range of interest.

For the turnoff and main sequence stars, only the merged ($V$, $B-V$) data of \citet{BE91, BO90} were adopted for stars within the survey areas. For a small subsample of stars far enough beyond the cluster core to lie outside the coverage of both $BV$ surveys, simple ($G$, $B_{p}-R_{p}$) transformations to ($V$, $B-V$) with linear terms in ($B_{p}-R_{p}$) were defined using only astrometric members of the cluster \citep{CA18} within the same magnitude range as the stars of interest at the turnoff. Unlike the multiple merger process for broad-band and intermediate-band photometry applied to the giants, the transformed $Gaia$ data were adopted only for the stars not found within the $BV$ CCD surveys.

A comprehensive discussion of the cluster reddening as of 2005 is given in ATAT. An early photometric study by \citet{K96} used $VI$ photometry to obtain a reddening estimate E$(V-I) = 0.10$, implying E$(B-V) = 0.077$ using E$(V-I)$ = 1.35*E$(B-V)$. This estimate depends in part on analysis of a nearby field RR Lyr which permits setting 0.08 as an upper limit for E$(B-V)$.  We note, however, that \citet{KA06b}, in their analysis of the eclipsing binary (EB), NV CMa, adopt the ATAT reddening value and an [Fe/H] similar to the value obtained therein to derive their EB mass, cluster distance, and age estimates. From the extended Str{\"o}mgren photometry of 
100 stars at the turnoff, ATAT derived  E$(B-V)$ = 0.055 $\pm$ 0.004 (sem). At the time, the only other estimate with reasonable precision was that of \citet{SC98} with $E(B-V)$ = 0.074, the maximum value along the line of sight in the direction of the cluster. Using the online tool\footnote{https://irsa.ipac.caltech.edu/applications/DUST/}, more recent corrections to the earlier reddening maps \citep{SC11} have reduced this value to a maximum reddening of $0.058^{+0.011}_{-0.005}$ in the direction of NGC 2243. Unfortunately, NGC 2243 sits at the edge of the map generated by \citet{GR19} and attempts to derive the trend along the line of sight fail to converge.

Finally, in Figure 1 we illustrate the location of the stars under discussion within the CMD of NGC 2243. Filled black circles and crosses are the members and nonmembers, respectively, from the combined samples of Tables 1 and 2. An asterisk notes the location of W2135.  Filled blue circles are the members from FR13, and the filled red circles are members from HP. 
Of the 10 identified nonmembers, only two near the top of the turnoff fall within the CMD trend defined by the cluster members. Four nonmembers populate the blue straggler region well above the turnoff and the other four occupy the red giant zone, again well off the track defined by the members.

A few stars considered members by FR13 and HP are not included in this plot nor in further analysis, due to ambiguous or missing astrometric information.  These include W3628 (referred to by 1035 by \citet{FR13}, using the second of two numbering schemes introduced by \citet{K96}), and W2512 (1106). An additional star, W1910, from the HP sample has also been excluded from further consideration. 
 
FR13 provide no additional information about the stars they identify as probable nonmembers based on radial velocities, but note that one star, Kaluzny 1410 (W2363), has an essentially featureless spectrum, so much so that no $V_{rad}$ estimate is possible.  This description bore a striking resemblance to the appearance of the Hydra spectrum for W1133; we were prompted to verify that like W1133, W2363 has been identified as variable star V3 by \citet{KA06}, an eclipsing binary NW CMa of W UMa class with a period of 0.356 days.  Although confirmed as an astrometric member by \citet{CA18}, the star is not included in the G3D sample, presumably because of a lack of a reliable $V_{rad}$ measurement. 

\section{Abundance Analysis: Hydra Spectra}

\subsection{Spectroscopic Processing With ROBOSPECT}

As described in \citet{LB15}, we have employed the automated equivalent width (EW) measuring software ROBOSPECT \citep{WH13}. Details \citep{LB15} are provided of our tests of the software and refinement of a linelist to exploit the iron and other heavy element lines included in the $\sim$400 \AA\ region surrounding the Li line. For the present investigation, we utilize the linelist refined in that paper, which includes 17 Fe, 3 Ni, 1 Ca and 1 Si lines.
 
In addition to the potential members of NGC 2243, daytime sky spectra were obtained in 2015 and 2016 using the identical Hydra fiber configuration and wavelength coverage. These spectra were analyzed using ROBOSPECT and used to reference the abundance from each line to the solar value by implementing minor adjustments, if necessary, to log $gf$ values for any line. This approach ensures that our abundances are relative to the solar value and solar spectrum rather than any adopted solar abundance value.
From all ROBOSPECT output, we discarded negative EW (ROBOSPECT's designation of emission lines) due to non-convergent fitting solutions, artifacts of cosmic ray removal, large noise spikes, or non-existent lines in the measured spectrum. We also excluded measured EW below 5 m\AA\ or above 150 m\AA. 

\subsection{$T_{\mathrm{eff}}$, Surface Gravity, and Microturbulent Velocity}


From measured EW, abundances are derived in the context of a stellar atmosphere model constructed with appropriate values of $T_{\mathrm{eff}}$, log $g$ and microturbulent velocity, $\xi$. The $T_{\mathrm{eff}}$ determination was approached in a manner to ensure consistency with previous spectroscopic studies by this group. For dwarfs, we continue to use the calibration defined in \citet{DE02}; for giants, the $T_{\mathrm{eff}}$-color-[Fe/H] calibration of \citet{RA05} was used. In both regimes, reddening values of E$(B-V) = 0.055$ were applied to $B-V$ colors, with a value of [Fe/H] $= -0.55$ adopted in the color-temperature calibration.

Estimates of the surface gravity were obtained for each likely member star in our spectroscopic sample by comparing its $V$, $B-V$ values to a Victoria-Regina isochrone \citep{VR06}(VR), constructed with the appropriate age, metallicity, and corrections for distance and reddening, as discussed in Section 4. Our approach to microturbulent velocity estimation follows the one used in \citet{LB15}.  For stars with surface gravity below 3.0, $\xi$ was estimated using a surface-gravity-based algorithm, $\xi = 2.0 - 0.2$ log $g$. For less evolved stars, we employ the $T_{\mathrm{eff}}$, log $g$ formulation developed by \citet{Br12}, based on comparison of spectroscopic and asteroseismic parameters for $Kepler$ stars.  

Model atmospheres were constructed for each program star using the grid of \citet{KU95} with input $T_{\mathrm{eff}}$, log $g$, and microturbulent velocity values. Each star's measured equivalent widths and model serve as input to the {\it abfind} routine of MOOG \citep{SN73} to produce individual [A/H] estimates for each successfully measured line in each star. With high confidence in the membership 
credentials for our spectroscopic sample, it makes sense to follow the statistical approach outlined in \citet{LB15}, including the consistent use of medians to estimate cluster values, as well as constructed Median Absolute Deviations (MAD) statistics to estimate the range for each estimated quantity. 

Figure 2 illustrates information presented in Table 4, including the [A/H] values obtained from each of the 22 lines in our linelist; error bars indicate the MAD statistic for abundance values derived from metal lines in as many as 29 stars (or as few as 14).  Clearly some lines produce noisier results than others, but we stress that use of median estimators lessens the impact of these lines.  The dashed horizontal line indicates an overall [Fe/H] value of $-0.55$, based on the following considerations.  Perhaps the most robust statistical estimation of the cluster's [Fe/H] is derived from the median of all 354 line measures, with a resulting median value of [Fe/H] $= -0.55$ and a MAD of 0.13 dex for all 354 separate estimators of [Fe/H]. An approximation of a standard deviation for our sample is given by 1.48*MAD, or 0.19.  

Consideration of the [Fe/H] values for each of the 17 Fe lines leads to a median value of [Fe/H] $= -0.53$ among the 17 estimations, MAD $= 0.10$ dex, estimated standard deviation 0.15. 
Taking the 29 separate [Fe/H] estimations for the sample stars, a median value of [Fe/H] = $-0.56$ is obtained, MAD $=0.12$ dex, $\sigma = 0.18$.  As a precaution, we recomputed this median using only stars with at least 8 successfully measured Fe lines, producing the same result.  

From a consistency standpoint, this result is encouraging. ATAT derived a weighted average of [Fe/H] = $-0.57$ $\pm$ 0.03 (sem) from
$m_1$ and $hk$ for 100 stars at the NGC 2243 turnoff, slightly lower than the typical result above. For NGC 2506, analyzed in the 
same manner as NGC 2243, \citet{AT18} derived a spectroscopic abundance of [Fe/H] = $-0.27$ $\pm$ 0.07 (sd) from 145 cluster giants and turnoff stars, only slightly higher than [Fe/H] = $-0.32$ $\pm$ 0.03 (sem) from 257 stars near the turnoff \citep{AT16}.

While individual [Ca/H], [Ni/H] and [Si/H] values, based on one to three lines for each star, are presented in Table 4, a more robust view is again supplied by median values for 
the 24-29 stars with successful line measurements.  From 24 stars, the median [Ca/H] $=-0.48 \pm 0.19$, where the quoted error indicates the MAD statistic, from 29 stars, [Ni/H] $=-0.61 \pm 0.06$ and, from 28 stars, median [Si/H] $=-0.44 \pm 0.11$.  Relative to the adopted cluster [Fe/H] $= -0.55$, these values represent [A/Fe] values of $0.07, -0.06$ and 0.10 for Ca, Ni and Si respectively, consistent within the uncertainties with little to no differential elemental enhancement in the cluster.  

\begin{figure}
\figurenum{2}
\plotone{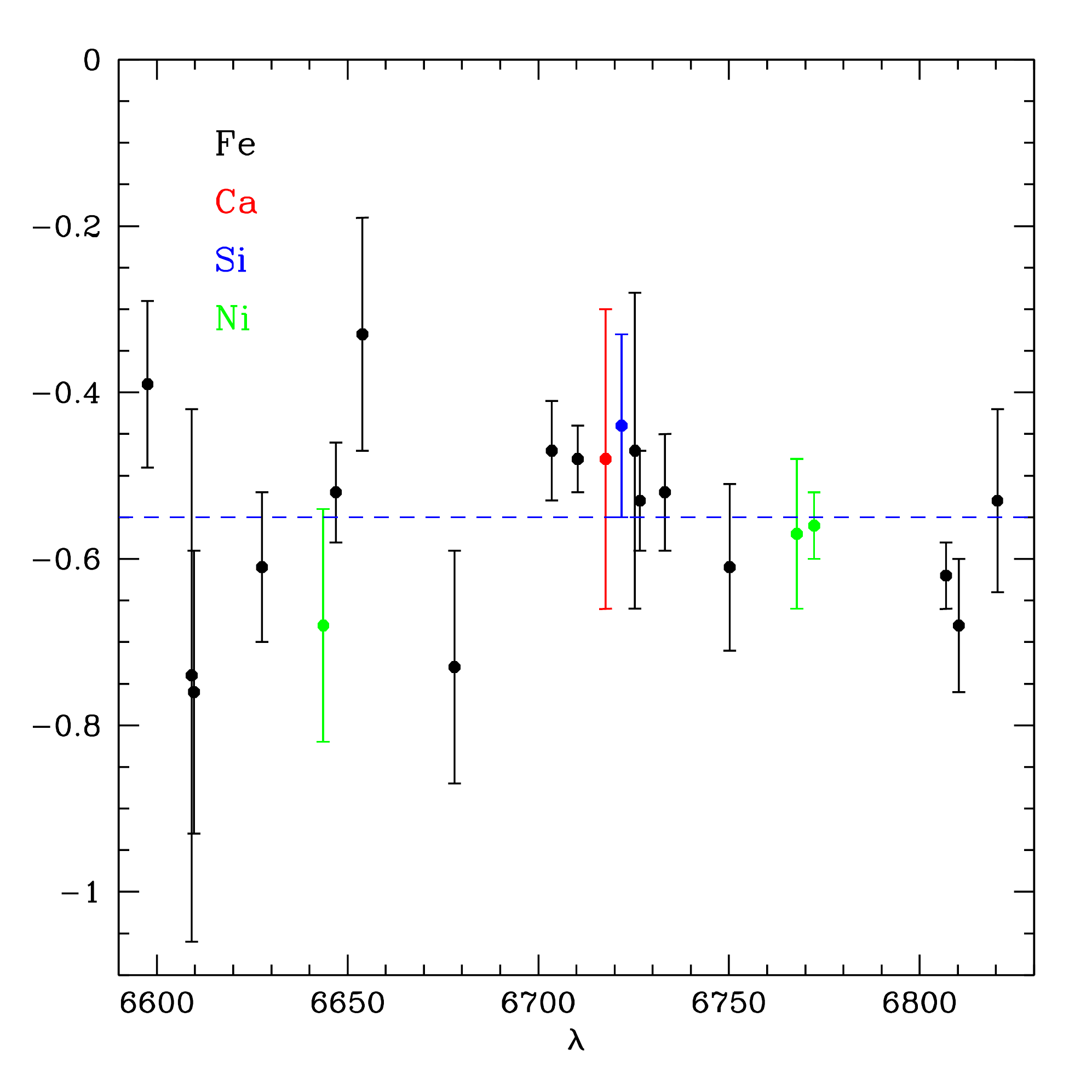}
\caption{Median abundances and median absolute deviation (MAD) statistics for each analyzed line in Hydra spectra.}
\end{figure}

Table 5 summarizes the input atmospheric parameters and spectroscopic results for each likely member star with a Hydra spectrum, with the exceptions of W2135, W1133 and W716 for reasons summarized above. Figure 3 illustrates the run of [Fe/H] values as a function of $T_{\mathrm{eff}}$ for which significant sensitivity might be a concern, and as a function of S/N for our sample.  No significant trends are discernible here, as was also the case for [Fe/H] with respect to log $g$ or $\xi$.

\subsection{Hydra Spectroscopic Uncertainties and Comparisons With Other Spectroscopic Work}
Probing the effects of incremented/decremented atmospheric parameters was particularly rewarding in the analysis of NGC 6819 Hydra spectroscopy by 
\citet{LB15} due to the large number of stars and the thoroughly populated CMD distribution, with an extensive error 
propagation map presented in Figure 5 of that paper. We probed a subset of our smaller sample to verify that similar trends are found in NGC 2243. Four stars in different parts of the CMD were reanalyzed with incremented/decremented atmospheric parameters to probe the sensitivity of [Fe/H] estimates to incorrect parameter choices, resulting in similar effects.  Although the results are not necessarily symmetric for increments/decrements in relevant quantities, we will conform to the convention presented in \citet{LB15} and summarize the consequences of adopted $T_{\mathrm{eff}}$ values too high by 100 K, surface gravities too small by 0.25, and microturbulent velocities too high by 0.25 km-s$^{-1}$.

For cooler stars, assignment of a $T_{\mathrm{eff}}$ too high by 100 K results in a derived [Fe/H] too high by $0.08$ dex, with a marginally smaller effect (+0.07) for stars near the turnoff.  It should be noted that a $\Delta$$T_{\mathrm{eff}}$ this large would imply a $B-V$ color or reddening error of 0.06 mag for cooler giants or 0.02 mag for stars nearer the turnoff. 
The abundance effect of log $g$ values assumed to be too small by 0.25 is only discernible for cooler giants, amounting to $\Delta$[Fe/H] $=-0.02$, ({\it i.e.}, lower log $g$ leads to lower abundance), again similar to results from \citet{LB15}.  Using microturbulent velocities too small by 0.25 km-s$^{-1}$ will produce a higher [Fe/H] although the effects are more $T_{\mathrm{eff}}$-dependent, smallest for turnoff stars (+0.02) to $\Delta$[Fe/H]$= +0.07$ dex for the cooler giants. 

As noted earlier, NGC 2243's place as one of the most metal-poor open clusters is fairly clear despite a historic range of [Fe/H] estimates that is distressingly large.  This history is summarized well by recent studies, including \citet{JA11} and \citet{Kov}.  It appears that differences between one study and another cannot always be simply explained as due to different $T_{\mathrm{eff}}$ scales although temperature is likely to have the dominant effect on [Fe/H] determination.  
One of the lower estimates, [Fe/H] = $-0.63$, was derived by \citet{HFC}, who employed $JHK$ photometry to derive $T_{\mathrm{eff}}$.  From four stars in common between our analysis set and theirs, we estimate that our $T_{\mathrm{eff}}$ values are hotter by $109 \pm 69$ K, which alone would account for our 0.08 dex higher estimate of [Fe/H].   

With respect to the parametric precepts of \citet{JA11}, we adopt $T_{\mathrm{eff}}$ only 35 K warmer and log $g$ values 0.3 dex  lower.  For giant stars, we would therefore expect these differences to generate approximately offsetting increments to the metallicity, yet the cluster metallicity from \citet{JA11} using Hydra spectra of higher resolution but lower S/N is [Fe/H] = $-0.42$, higher than our estimate by more than 0.10 dex.

\citet{MA17} provide detailed $T_{\mathrm{eff}}$, log $g$ and $\xi$ estimates for 13 cool stars in common with the present study, with a resulting cluster average [Fe/H] of $-0.35$,  $\sim$0.2 dex higher than our estimate.  Based on $BV$ photometry, we would assign $T_{\mathrm{eff}}$ 60 K cooler, log $g$ values 0.1 dex smaller and $\xi$ values 0.13 km/sec larger, which would work in the same sense to lower the \citet{MA17} [Fe/H] estimate by no more than 0.07 dex, placing it in the [Fe/H] = $-0.4$ to $-0.5$ range. 

Sufficient $BV$ photometry exists for stars studied by FR13 to estimate atmospheric parameters for their sample stars in a manner consistent with precepts followed in this study.  As noted by FR13, their temperature scale is relatively hot.  Differences between their adopted $T_{\mathrm{eff}}$ and values assigned by color-temperature relations from $BV$ photometry range from 150 K for stars below the clump to 250 K or higher for stars near the turnoff.  Despite a prediction that lowering their temperature scale would result in lower [Fe/H] and [Li/H] values by $\geq 0.15$ dex, FR13 derive [Fe/H] = $-0.54$, virtually identical to the median value derived from EW measures in the current study.

Accounting for differences between spectroscopic analyses may require looking beyond the most traceable effects ({\it e.g.} differing temperature scales) to causes more difficult to probe.  These may include atmospheric models, linelists and adopted log $gf$ values in addition to operational differences, {\it e.g.}, continuum level estimation and line measurement techniques. Alternatively, it may be useful to consider methods that train and utilize neural networks to analyze spectra in a holistic manner.

We previously developed and used such a neural network classification approach called ANNA \citep{LB15, LB17, LB18} in similar contexts, especially NGC 2506 and NGC 6819 \citep{AT18, DE19}. Data from NGC 2506 successfully analyzed using ANNA were obtained in the course of the same runs as our Hydra spectra in NGC 2243, so the same training parameters developed could be applied to NGC 2243 Hydra spectra.  This application is not obviously straightforward because the ANNA training set was developed using Kurucz atmospheric models more metal-rich than [Fe/H] = $-0.50$, making extrapolation to lower [Fe/H] unreliable.  We restrict our reference to ANNA results for the cooler, evolved stars in NGC 2243 for which the median [Fe/H] value for 16 stars is $-0.42  \pm 0.03$, keeping in mind that this value is a likely upper limit to the true cluster metallicity.

\citet{Kov} have also developed a neural-network approach to parameter estimation, using both LTE and non-LTE atmospheric models.  As with ANNA, ``the Payne'' code first trains on a large set of synthesized spectra then performs fits. With a training set built upon atmospheric models that include considerably lower [Fe/H] values than ANNA, ``the Payne'' is well situated to provide parameter estimates for globular clusters as well as open clusters, as described in \citet{Kov} where NGC 2243 and a more metal-rich open cluster, NGC 3532, are discussed.  82 likely members of NGC 2243 were selected using radial-velocity and $Gaia$ proper-motion criteria. An intra-cluster spread of 0.07 dex in [Fe/H] is estimated for NGC 2243.  From the limited specific data provided for NGC 2243 stars, it appears that their derived $T_{\mathrm{eff}}$ for a turnoff star is approximately 150 K hotter than our $BV$ photometry-based values. The LTE result from \citet{Kov}, probably the most appropriate comparison to our LTE-based results, is [Fe/H] $= -0.57 \pm 0.11$, where the quoted error refers to a mean systematic error; the NLTE estimate is 0.05 dex higher.
  
Possible $\alpha$-element enhancement in NGC 2243 has been a topic of interest due to its location in the outer disk of the Milky Way. \citet{GC94} analyzed two giants to arrive at [Fe/H] of $= -0.48 \pm 0.15$  for the cluster, paying additional attention to a variety of other elements, including $\alpha$-elements, concluding that a scaled solar abundance pattern was neither excluded nor strongly favored, with typical enhancements [$\alpha$/Fe] of 0.07.   More recently, \citet{JA11} find elemental enhancements [X/Fe] $\leq 0.15$, consistent with scaled solar abundances.  In particular, they cite strong evidence of solar [O/Fe] for one of the giants in NGC 2243.

Similarly, \citet{Kov} cite only modest evidence for $\alpha$-element enhancement, +0.20 $\pm$ 0.22 for [Ti/Fe], while FR13 again find [Fe/H] = -0.54 $\pm$ 0.10, with modest or null indications of enhancement for Ca or Si.

\begin{deluxetable}{ccccc}
\tablenum{4}
\tablecaption{Characteristics and Results for Individual Metal Lines}
\tablewidth{700pt}
\tablehead{
\colhead{Elem.} & \colhead{$\lambda$} & \colhead{[A/H]} & \colhead{Number} & \colhead{MAD} \\
\colhead{} & \colhead{\AA} & \colhead{} & \colhead{of Stars} & \colhead{} }
\startdata
Fe & 6597.56 & -0.39 & 25 & 0.10  \\
Fe & 6609.11 & -0.74 & 23 & 0.32  \\
Fe & 6609.68 & -0.76 & 20 & 0.17 \\
Fe & 6627.54 &  -0.61 & 22 & 0.09 \\
Fe & 6646.93 &  -0.52 & 14 & 0.06 \\
Fe & 6653.91 &  -0.33 & 14 & 0.14 \\
Fe & 6677.99 &  -0.73 & 20 & 0.14 \\
Fe & 6703.57 & -0.47 & 24 & 0.06  \\
Fe & 6710.32 & -0.48 & 18 & 0.04 \\
Fe & 6725.36 & -0.47 & 17 & 0.19 \\
Fe & 6726.67 & -0.53 & 25 & 0.06 \\
Fe & 6733.15 & -0.52 & 20 & 0.07 \\
Fe & 6750.15 & -0.61 & 28 & 0.10 \\
Fe & 6806.86 & -0.62 & 21 & 0.04 \\
Fe & 6810.27 & -0.68 & 24 & 0.08 \\
Fe & 6820.37 & -0.53 & 23 & 0.11 \\
Fe & 6837.01 & -0.44 & 16 & 0.19 \\
Ca & 6717.68 & -0.48 & 25 & 0.18 \\
Si & 6721.85 & -0.44 & 28 & 0.11 \\
Ni & 6643.63 & -0.68 & 28 & 0.14 \\
Ni & 6767.77 & -0.57 & 29 & 0.09 \\
Ni & 6772.31 & -0.56 & 27 & 0.04 \\
\enddata
\end{deluxetable} 

\begin{figure}
\figurenum{3}
\plotone{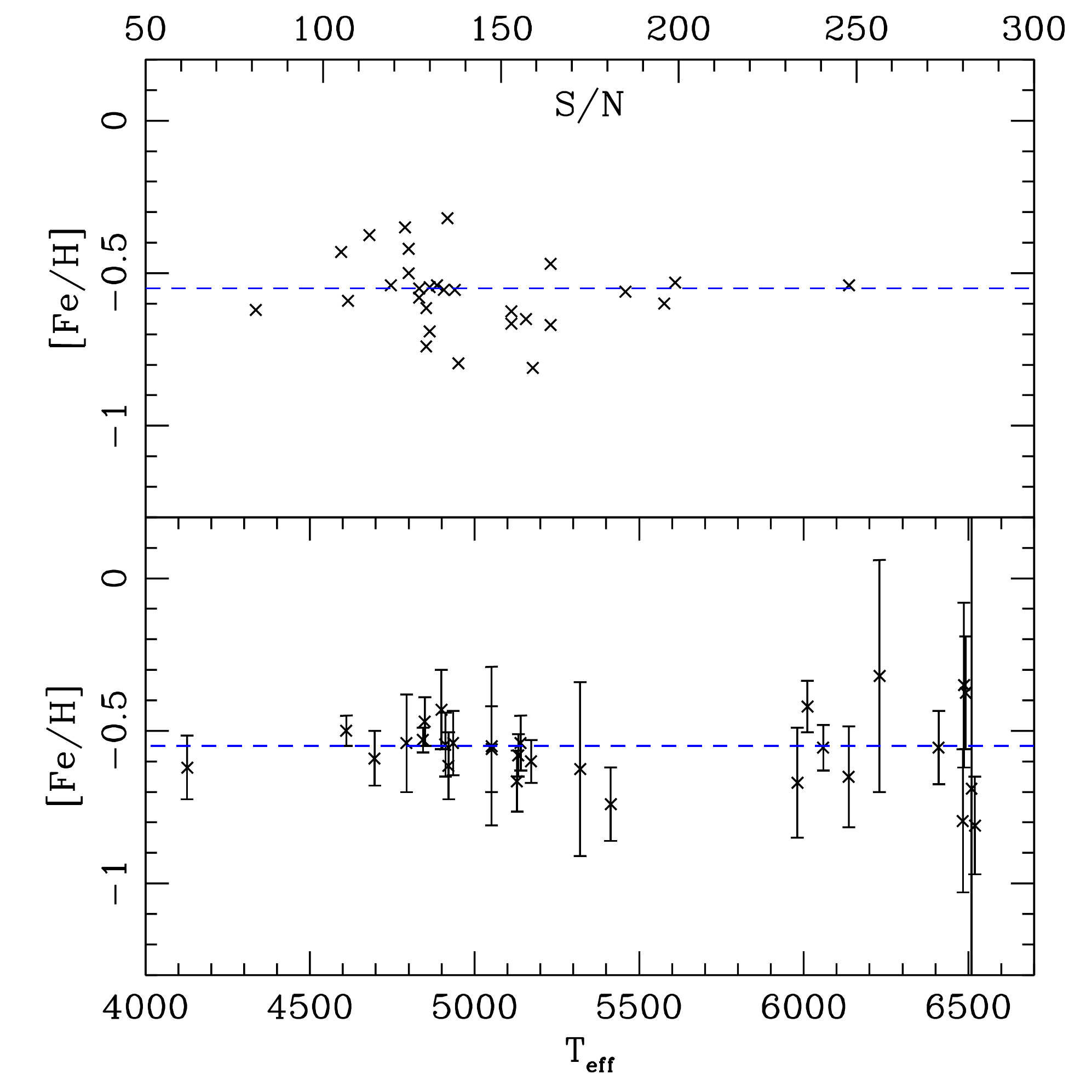}
\caption{[Fe/H] as a function of $T_{\mathrm{eff}}$ and S/N.}
\end{figure}

\floattable
\begin{deluxetable}{rrrrrrrrrr}
\tablenum{5}
\tablecaption{Individual Abundance Information for NGC 2243 Stars, Hydra Sample}
\tablewidth{0pt}
\tablehead{
\colhead{W ID} & \colhead{T$_{eff}$} & \colhead{log $g$} & \colhead{$\xi$} & \colhead{[Fe/H]} &
\colhead{MAD} & \colhead{Num.} & \colhead{[Ca/H]} & \colhead{[Ni/H]} & \colhead{[Si/H]}   } 
\startdata
519 & 6199 & 3.75 & 1.72 & -0.62 & 0.17 & 10 & -0.65 & -0.65 & \nodata \\
611 & 5052 & 3.25 & 1.34 & -0.56 & 0.14 & 17 & -0.40 & -0.55 & -0.35\\
873 & 6055 & 3.75 & 1.61 & -0.62 & 0.19 & 11 & -0.46 & -0.55 & -0.47\\
1044 & 6337 & 3.80 & 1.82 & -0.26 & 0.37 & 10 & -0.74 & -0.62 & -0.55\\
1230 & 5321 & 3.60 & 1.28 & -0.67 & 0.24 & 11 & -0.42 & -0.74 & -0.39\\
1263 & 5172 & 3.20 & 1.38 & -0.60 & 0.07 & 15 & -0.43 & -0.63 & -0.71\\
1266 & 6089 & 3.75 & 1.64 & -0.37 & 0.10 & 10 & -0.57 & -0.41 & -0.21\\
1271 & 4935 & 2.50 & 1.50 & -0.55 & 0.12 & 15 & -0.31 & -0.60 & -0.40\\
1294 & 5052 & 3.25 & 1.34 & -0.55 & 0.26 & 17 & -0.48 & -0.66 & -0.53\\
1313 & 4610 & 2.00 & 1.60 & -0.50 & 0.05 & 15 & \nodata  & -0.53 & -0.34\\
1421 & 5129 & 3.40 & 1.30 & -0.66 & 0.10 & 14 & -0.66 & -0.72 & -0.12\\
1436 & 6396 & 3.85 & 1.85 & -0.57 & 0.12 & 8 & -0.92 & -0.77 & -0.66\\
1467 & 4911 & 2.50 & 1.50 & -0.52 & 0.11 & 15 & -0.21 & -0.49 & -0.67\\
1696 & 4843 & 2.65 & 1.47 & -0.53 & 0.04 & 16 & -0.28 & -0.56 & -0.35\\
1738 & 4921 & 2.50 & 1.50 & -0.66 & 0.15 & 15 & -0.12 & -0.51 & -0.40\\
1847 & 4849 & 2.65 & 1.47 & -0.47 & 0.08 & 16 & -0.25 & -0.61 & -0.25\\
1871 & 6387 & 3.85 & 1.84 & -0.89 & 0.15 & 5 & -0.70 & -0.83 & -0.55\\
1995 & 5133 & 3.40 & 1.30 & -0.58 & 0.07 & 14 & -0.37 & -0.61 & -0.29\\
2003 & 6446 & 3.85 & 1.90 & -0.82 & 0.23 & 4 & -0.83 & -0.78 & -0.96\\
2098 & 5414 & 3.60 & 1.31 & -0.74 & 0.12 & 13 & -0.59 & -0.76 & -0.44\\
2394 & 6128 & 3.75 & 1.67 & -0.51 & 0.06 & 6 & -0.65 & -0.56 & -0.49\\
2410 & 4899 & 2.50 & 1.50 & -0.43 & 0.12 & 16 & -0.09 & -0.47 & -0.45\\
2434 & 5140 & 2.60 & 1.48 & -0.54 & 0.09 & 17 & -0.24 & -0.55 & -0.33\\
2676 & 6506 & 3.95 & 1.92 & -0.33 & 0.28 & 8 & -0.73 & -0.71 & -0.18\\
2696 & 6529 & 3.95 & 1.94 & -0.68 & NA & 1 & -0.77 & -0.62 & -0.50\\
2908 & 6510 & 3.95 & 1.92 & -0.36 & 0.19 & 8 & -0.66 & -0.60 & -0.87\\
3618 & 4696 & 2.35 & 1.53 & -0.60 & 0.10 & 16 & \nodata  & -0.74 & -0.56\\
3633 & 4127 & 1.20 & 1.76 & -0.64 & 0.07 & 15 & \nodata  & -0.71 & -0.80\\
3728 & 4793 & 2.25 & 1.55 & -0.57 & 0.15 & 16 & \nodata  & -0.55 & -0.31\\
\enddata
\tablecomments{No equivalent width analysis for W2135 was carried out; corresponding atmospheric model parameters would be 
4652 K, 2.35 and 1.53 km/sec.}
\end{deluxetable}

\section{The CMD: Cluster Age and Distance}
To place the stellar population within NGC 2243 and the cluster itself in the appropriate context for evaluation and interpretation of the Li patterns described below, we show in Figure 4 the CMD of well-defined cluster members relative to an appropriate set of VR isochrones. The filled circles include all member cluster giants with $B-V > 0.70$; the open circles are core cluster members with $Gaia$ probabilities $\ge$ 0.7 and CCD $BV$ photometry while the asterisk again notes W2135.  The compilation of the $BV$ photometry is described in Section 2.4. The adopted isochrones have [Fe/H] = $-0.52$ and are scaled-solar models. While slightly higher than the spectroscopically derived range of [Fe/H] = $-0.53$ to $-0.56$ (Section 3.2), the minor offset of the isochrones can account, in part, for the possibility of a minor $\alpha$-element enhancement. For consistency with the standard adopted in our previous cluster analyses using VR isochrones, we first zero the isochrones by requiring that a star of solar mass and metallicity  at an age of 4.6 Gyr have $M_{V}$ = 4.84 and $B-V$ = 0.65, leading to minor adjustments, $\Delta$$V$ = 0.02 and $\Delta$$(B-V)$ = +0.013 mag. The assumed cluster reddening is E$(B-V)$ = 0.055 (Section 2.5). The isochrones have been shifted by an apparent distance modulus of 13.20 and have ages of 3.4, 3.6, and 3.8 Gyr.

The average parallaxes for the stars plotted in Figure 4 yield an apparent modulus of $(m-M)$ = 13.55 with an appropriate contribution from the adopted reddening.  
However, this does not account for the often discussed likelihood of a zero-point offset of -0.05 $\pm$ 0.03 mas in the $Gaia$ parallaxes, especially for stars at greater distance (see, {\it e.g.} \citet{CA18, RI18, ST18, ZI19}). Applying a typical correction of +0.05 mas to the cluster $Gaia$ DR2 parallax generates $(m-M)$ = 13.1, clearly in much better agreement with the isochrone fit. 

While the isochronal giant branches are an excellent match to the observational trends, 
the colors of the turnoff stars extend across the full color range of the isochrones, showing more structure than implied by the models. 
The luminosities of the subgiant stars also appear to lie above the isochrones, which might reflect an age on the lower side of our proposed range.
As a compromise, we will adopt 3.6 $\pm$ 0.2 Gyr as the cluster age and use that isochrone to transfer from the observational to the theoretical plane and to delineate the stellar masses for the stars within our Li sample. 
 
\begin{figure}
\figurenum{4}
\plotone{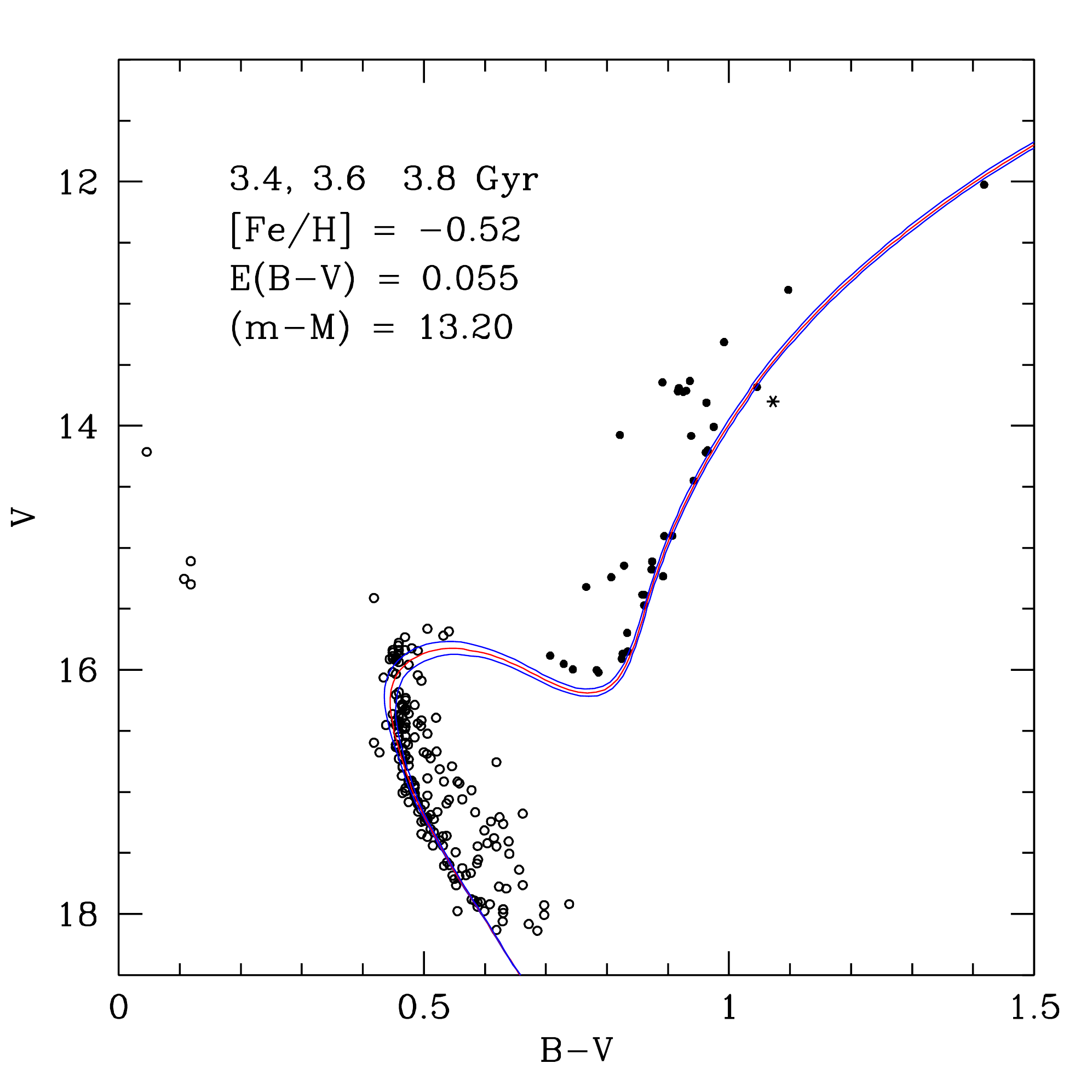}
\caption{CMD comparison to the isochrones of \citet{VR06}. Isochrones have ages of 3.4, 3.6, and 3.8 Gyr. Filled circles include all cluster members with $B-V > 0.7$. Open circles are highly probable members from the cluster core with CCD $B-V$ photometry while an asterisk notes the location of W2135.}
\end{figure}

\section{Lithium Abundances and Trends}

\subsection{Spectrum Synthesis and Data Merger}
We have utilized spectrum synthesis techniques to estimate A(Li) for all high S/N spectra in our Hydra sample, augmented by available archive spectra from the VLT; these stars have summarized in Tables 1 and 2 respectively.  
For this purpose, we used the {\it synth} mode of MOOG \citep{SN73}, referencing the spectra to model atmospheres appropriate to each star.  As described in Section 3.2, $T_{\mathrm{eff}}$ are derived in an internally consistent manner from homogeneous $B-V$ colors using two color-temperature relations adopted in previous investigations by this group. Interactive estimation of A(Li) in the synthesis process also provides an opportunity to verify that the adopted $T_{\mathrm{eff}}$ for each star is appropriate, based on simultaneous agreement of the several temperature-dependent lines near 6710 \AA. 

Table 6 summarizes the results of synthesis on all spectra, with consensus A(Li) estimates that summarize results from Hydra, Giraffe and UVES spectra.  S/N values for the Hydra and VLT samples may be found in Tables 1 and 2. While a standard for S/N values generally $\geq 100$ is desirable, a few exceptions were made for VLT spectra to deliberately expand the very small overlap with the previous Li abundance samples of FR13 and HP. Adopted atmospheric model parameters are found in Tables 5 and 2 for the two spectroscopic samples.
The consensus or overall A(Li) value was arrived at with the following general considerations: spectra with higher S/N were favored, with higher resolution dominating only if S/N values are comparable.  All other factors being comparable, results from our own Hydra spectra were favored. Of 14 Hydra stars for which VLT spectra were also available, 13 fell within the same category for both sources, {\it i.e.} classed as detections or upper limits in both cases. Only W3618 was switched from an upper limit of A(Li) = 0.65 using Hydra spectra to a detection of A(Li) = 0.65 from VLT. 

Comparison of these Li abundances with prior work is complicated by differences of methodology and temperature scales, with the expectation that the latter would provide the dominant influence.  FR13 make primary use of $VI$ photometry from \citet{K96}, coupled with their fairly high reddening correction, equivalent to almost E$(B-V) = 0.08$.  FR13 note in their paper that a prior study by HP employs a lower reddening value with consequently lower temperatures.  Figure 6 of FR13 illustrates both samples in a magnitude, $T_{\mathrm{eff}}$ plane with a substantial temperature offset between the two samples amounting to 200 K and resulting in Li abundances lower by 0.15 dex in HP.  

Figure 1 illustrated our sample (filled black points), as well as those analyzed by FR13 (blue points) and HP (red points); the photometry is the $BV$ photometry described in Section 2.5. Clearly no intrinsic $T_{\mathrm{eff}}$ offset, as defined by $B-V$, between the samples is evident.  It remains important, therefore, to 
investigate the $T_{\mathrm{eff}}$ differentials before discussing all available Li abundances in a homogenous manner.  

Having found or synthesized colors for all of HP's and FR13's sample, we estimated $T_{\mathrm{eff}}$ values for all stars based on color-temperature relations discussed above.  FR13's use of a higher reddening value than used in the present study implies $T_{\mathrm{eff}}$ for near turnoff stars that are $\geq 250$ K higher in their analysis, while cooler giants are taken to be $\geq 150$ K hotter than our assessment would be. FR13 themselves estimate that temperatures higher by 200 K imply A(Li) values higher by 0.15 dex.  We sought to validate this estimate in two ways.  First, there are a handful of stars with abundances based on our synthesis analysis that overlap with the FR13 sample, none of them from our original Hydra sample but from the augmented archival VLT spectra.  Three stars from Table 2 with detections overlap with FR13 detections, W1261, W1686 and W2648.  In the case of W1686, our analysis provides A(Li) that is 0.24 dex higher than FR13; conversely, our analysis yields a Li abundance for W2648 that is 0.36 lower. Our estimated A(Li) for W1261 is only 0.04 dex smaller than that of FR13.  No direct overlap of samples between the results presented in Tables 1 and 2 with that of HP exists at all.  

Another way to estimate the effect of increments in $T_{\mathrm{eff}}$ is to repeat the synthesis analysis on our available spectra using models with higher temperatures.  Test cases suggest that use of models with temperatures higher by 200 K for cooler stars and 300 K for turnoff stars would produce Li abundances values higher by 0.2 dex.  It is unclear whether other differences in methodology (continuum placement, use of equivalent widths, a higher assumed overall metallicity) between our results and those from other surveys mask or negate some of this offset.  Since the direct comparison of our A(Li) values for two stars in common with FR13 is indeterminate, we elect to present Li abundances from FR13 and HP with no adjustment in the following discussion. 

\floattable
\begin{deluxetable}{rhhhrrrrrrr}
\tablenum{6}
\tablecaption{A(Li) Derived from Synthesis}
\tablewidth{0pt}
\tablehead{
\colhead{ID} & \nocolhead{T$_{eff}$} & \nocolhead{log $g$} & \nocolhead{$\xi$} &
\colhead{Adopted} & \multicolumn{2}{c}{Hydra} & \multicolumn{2}{c}{Giraffe} & \multicolumn{2}{c}{UVES} \\
\colhead{} & \colhead{} & \colhead{} & \colhead{} & \colhead{A(Li)} & 
\colhead{A(Li)} & \colhead{EW} & \colhead{A(Li)} & \colhead{EW} & \colhead{A(Li)} & \colhead{EW\tablenotemark{a}} }

\startdata
239 & 4990 & 2.5 & 1.5 &  $\leq 0.6$  &  \nodata    &  \nodata    &  $\leq 0.6$ & 7 &  $\leq 0.7$  & 1.8 \\
365 & 4824 & 2.5 & 1.5 & 0.75 &  \nodata    &  \nodata    &  \nodata    &  \nodata    & 0.75 & 16   \\
519 & 6199 & 3.75 & 1.72 & 2.9 & 2.9 & 82.2 &  \nodata    &  \nodata    &  \nodata  &  \nodata      \\
611 & 5052 & 3.25 & 1.34 & 1.1 & 1.1 & 31.5 & 1.0 & 26 &  \nodata  &  \nodata    \\ 
873 & 6055 & 3.75 & 1.61 & 2.45 & 2.45 & 44.2 &  \nodata    &  \nodata    &  \nodata  &  \nodata    \\ 
910 & 4939 & 2.5 & 1.5 &  $\leq 0.1$  &  \nodata    &  \nodata    &  $\leq 0.6$ & 12.8 &  $\leq 0.1$  & 3.1 \\ 
1044 & 6337 & 3.8 & 1.82 & 2.6 & 2.6 & 39 &  \nodata    &  \nodata    &  \nodata  &  \nodata     \\
1230 & 5321 & 3.6 & 1.28 &  $\leq 1.1$  &  $\leq 1.1$ & 7 &  \nodata    &  \nodata    &  \nodata  &  \nodata    \\ 
1261 & 5027 & 3.1 & 1.39 &  1.2  &  \nodata    &  \nodata    &  1.2 & 16 &  \nodata  &  \nodata      \\
1263 & 5172 & 3.2 & 1.38 &  $\leq 1.1$  &  $\leq 1.1$ & 13 &  $\leq 1.0$   & 12 &  \nodata  &  \nodata     \\
1266 & 6089 & 3.75 & 1.64 & 2.4 & 2.4 & 49.4 & 2.4 & 42.3 &  \nodata  &  \nodata     \\
1271 & 4935 & 2.5 & 1.5 &  $\leq 0.6$  &  $\leq 0.6$ & 10.5 &  $\leq 0.6$ & 15 &  $\leq 0.4$  & 3.5 \\ 
1294 & 5052 & 3.25 & 1.34 &  $\leq 0.5$  &  $\leq 0.5$ & 7.9 &  $\leq 0.8$ & 5 &  \nodata  &  \nodata      \\
1313 & 4610 & 2 & 1.6 &  $\leq 0.0$  &  $\leq 0$   & 13.5 &  \nodata    &  \nodata    &  $\leq -0.2$  & 2.6   \\
1421 & 5129 & 3.4 & 1.3 &  $\leq 0.6$  &  $\leq0.6$  & 4.2 &  \nodata    &  \nodata    &  \nodata  &  \nodata     \\
1436 & 6396 & 3.85 & 1.85 & 2.4 & 2.4 & 19 &  \nodata    &  \nodata    &  \nodata  &  \nodata      \\
1467 & 4911 & 2.5 & 1.5 &  $\leq 0.1$  &  $\leq0.1$  & 11.5 &  $\leq 0.6$ & 20.6 &  $\leq 0.2$  & 5 \\ 
1679 & 5025 & 3.1 & 1.39 &  $\leq 1.2$  &  \nodata    &  \nodata    &  $\leq 1.2$ & 9 &  \nodata  &  \nodata     \\
1686 & 6324 & 3.8 & 1.77 & 2.85 &  \nodata    &  \nodata    & 2.85 & 49 &  \nodata  &  \nodata     \\
1696 & 4843 & 2.65 & 1.47 &  $\leq 0.6$  &  $\leq 0.6$ & 11.6 &  \nodata   &  \nodata   &  $\leq 0.6$  & 16  \\
1738 & 4921 & 2.5 & 1.5 &  $\leq 0.3$  &  $\leq 0.3$ & 8.8 &  \nodata   &  \nodata   &  \nodata  &  \nodata     \\
1847 & 4849 & 2.65 & 1.47 & 0.75 & 0.75 & 26.7 &  \nodata   &  \nodata   & 0.95 & 16  \\
1871 & 6387 & 3.85 & 1.84 & 2.3 & 2.3 & 20.4 &  \nodata   &  \nodata   &  \nodata  &  \nodata     \\
1923 & 6538 & 3.9 & 1.96 &  $\leq 2.1$  &  \nodata    &  \nodata    & $\leq 2.1$ & 3.3 &  \nodata  &  \nodata     \\
1995 & 5133 & 3.4 & 1.3 &  $\leq 0.8$  &  $\leq 0.8$ & 7.8 &  \nodata   &  \nodata   &  \nodata  &  \nodata     \\
2003 & 6446 & 3.85 & 1.9 & 2.6 & 2.6 & 33.4 &  \nodata   &  \nodata   &  \nodata  &  \nodata     \\
2098 & 5414 & 3.6 & 1.31 &  $\leq 1.4$  &  $\leq 1.4$ & 7.2 &  \nodata   &  \nodata   &  \nodata  &  \nodata    \\ 
2135 & 4652 & 2.35 & 1.53 & 1.0 & 1.0 & 42.6 & 1.0 & 43 & 1.05 & 78  \\
2195 & 5267 & 3.2 & 1.41 &  $\leq 1.2$  &  \nodata    &  \nodata    & $\leq 1.2$ & 5 &  \nodata  &  \nodata    \\ 
2394 & 6128 & 3.75 & 1.67 & 3.1 & 3.1 & 112 &  \nodata   &  \nodata   &  \nodata  &  \nodata     \\
2410 & 4899 & 2.5 & 1.5 &  $\leq 0.3$  &  $\leq 0.5$ & 17.3 &  \nodata   &  \nodata   &  $\leq 0.3$  & 4.5 \\ 
2434 & 5140 & 2.6 & 1.48 &  $\leq 0.7$  &  $\leq 0.7$ & 11.2 &  \nodata   &  \nodata   &  \nodata  &  \nodata     \\
2536 & 4987 & 3.1 & 1.38 &  $\leq 0.7$  &  \nodata    &  \nodata    &  $\leq 0.7$ & 9 &  \nodata  &  \nodata     \\
2613 & 5124 & 3.1 & 1.41 & 1.25 &  \nodata    &  \nodata    & 1.25 & 28 &  \nodata  &  \nodata     \\
2648 & 4888 & 2.75 & 1.5 & 1.15 &  \nodata    &  \nodata    &  \nodata   &  \nodata   & 1.15 & 39  \\
2676 & 6506 & 3.95 & 1.92 & 2.0 & 2.0 & 17.2 &  \nodata    &  \nodata    &  \nodata   &  \nodata     \\
2696 & 6529 & 3.95 & 1.94 & 2.4 & 2.4 & 22.4 &  \nodata   &  \nodata   &  \nodata  &  \nodata     \\
2908 & 6510 & 3.95 & 1.92 & 2.1 & 2.1 & 17 &  \nodata   &  \nodata   &  \nodata  &  \nodata    \\
3618 & 4696 & 2.35 & 1.53 & 0.65 & $\leq 0.65$ & 33.3 &  \nodata   &  \nodata   & 0.65 & 15  \\
3633 & 4127 & 1.2 & 1.76 &  $\leq -0.9$ & $\leq -0.9$ & 33.9 &  \nodata   &  \nodata   &  $\leq -0.4$  & 12  \\
3728 & 4793 & 2.25 & 1.55 &  $\leq 0.4$  & $\leq 0.4$  & 15.2 &  $\leq 0.0$  & 14.5 &  $\leq 0.1$ & 4.7  \\
\enddata
\tablenotetext{a}{Measured equivalent widths for UVES spectra refer only to Li line at 6707.78 \AA. EWs listed for HYDRA and Giraffe spectra include contribution from nearby Fe line.}
\end{deluxetable}

\subsection{Li Trends: The Turnoff Region}

Figure 5 illustrates the run of A(Li) values with respect to $V$ magnitude and $B-V$ color for all samples, with inverted triangles noting the location of upper limit values.  A number of features are immediately apparent. First, in the CMD region where the FR13 data, the stars of Tables 1 and 2, and the A(Li) results of HP overlap, {\it i.e.} at the top of the turnoff with $V < 16.0$, the FR13 sample (blue points) does not separate toward a higher A(Li) from the current sample (black points) or HP (red points), as one might expect given the higher temperature scale used by FR13. The largest detected Li abundance in the FR13 analysis is A(Li) = 2.9, 0.2 dex lower than the highest abundance derived for our sample. The probability of attaining a higher limiting A(Li) is undoubtedly enhanced by a sample of 12 stars from the current study relative to 4 from FR13, one of the reasons for boosting the population of stars observed within this critical evolutionary phase. Second, in contrast, the generally fainter stars explored in great depth by FR13 map out the Li-dip extraordinarily well, particularly at the faint end, but the addition of 5 members near $V$ = 16.0 from the current study and HP combine to illustrate the strikingly steep high-mass edge of the Li-dip, hereinafter referred to as the wall, a feature that emerges in all well-populated cluster samples older than the Hyades \citep{TW20} where stars entering the subgiant branch have masses outside the range of the Li-dip. The addition of these stars shifts the start of the Li-dip approximately 0.2 mag fainter than allowed from the FR13 data alone. Whether the limit could be pushed another 0.1 mag fainter still requires an even more expanded sample and cannot be answered at present. Third, the cool edge of the Li-dip where measurable detections of Li, rather than upper limits, recur is starkly apparent near $V$ = 16.95. While one could argue that the spread in A(Li) at this magnitude level  runs from the upper limit above the Li-dip to the lowest measured value within the Li-dip, FR13 has emphasized that the stars below $V$ = 16.95 with only detection limits are the likely product of spectra with low S/N and unlikely to be indicative of the true Li abundance among the fainter stars. Unfortunately, FR13 don't supply S/N data for individual stars, so the role of this parameter in setting the scatter among the fainter dwarfs remains ambiguous. Fourth, among the stars above the Li-dip ($V < 16.05$), A(Li) covers the range from $\sim$3.1 to an upper limit near 1.75. Unlike the scatter in A(Li) at the faint end, the S/N of the spectra among the majority of stars in this CMD region is high enough that the scatter is reflective of a real scatter in abundance. We will return to this key observation when discussing evolution from the main sequence through the giant branch in the following section. 

\begin{figure}
\figurenum{5}
\plotone{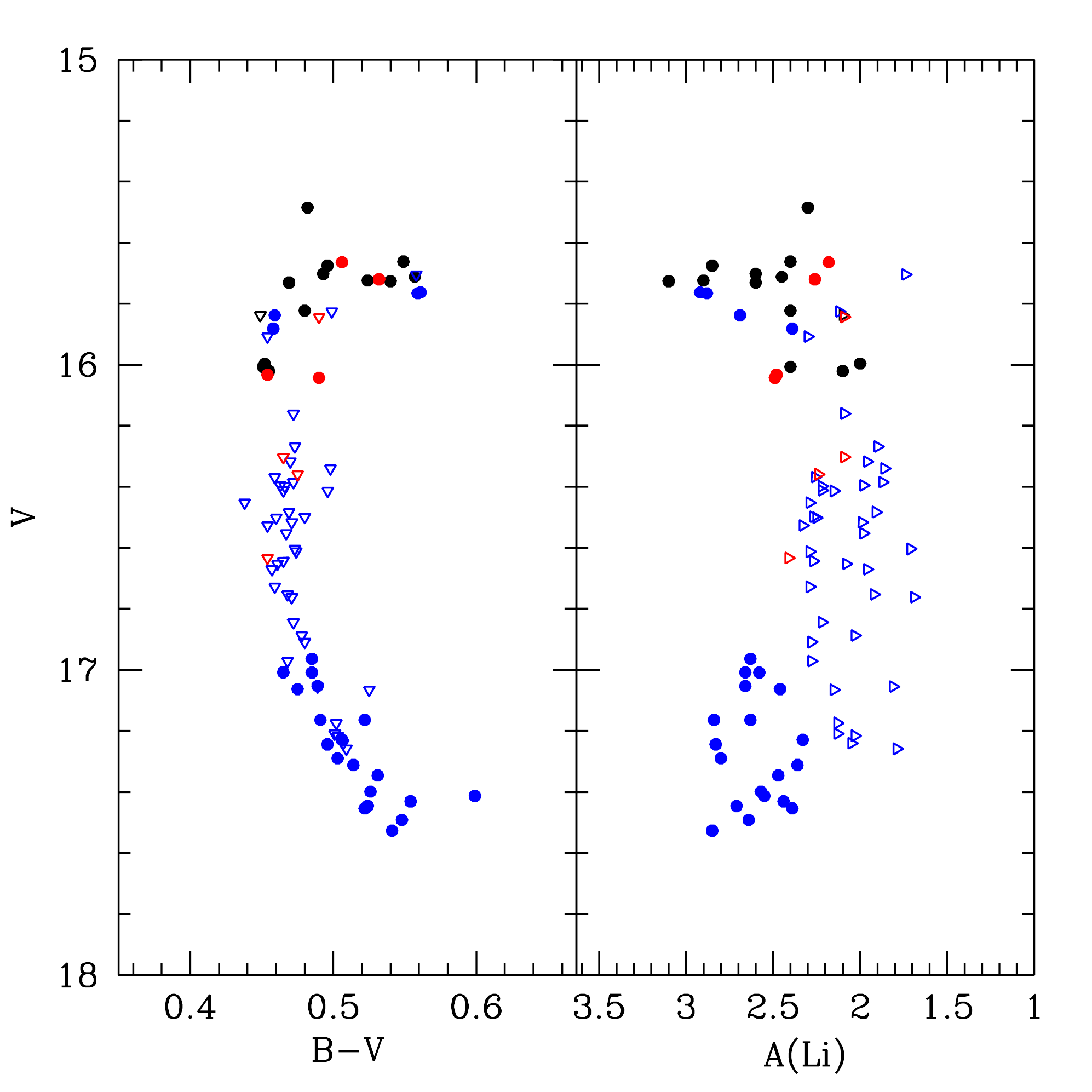}
\caption{A(Li) estimates for stars at the CMD turnoff. Symbol colors have the same meaning as in Figure 1. Filled circles are Li detections while triangles denote upper limits to A(Li).}
\end{figure}

Using the isochrones of Figure 4 and an age of 3.6 Gyr, we can translate the Li trend with $V$ at the turnoff into Li versus mass, as shown in Figure 6. Stars from the subgiant branch beyond $B-V$ = 0.7 and all red giants have been excluded from the plot because the mass range among these stars is too small to allow differentiation on this scale. Taking the discrete nature of the mass estimates into account and reiterating the greater uncertainty in the Li estimates at the lower mass end of the distribution, a plausible mass range encompassing the Li-dip for NGC 2243 runs from 1.09 $\pm$ 0.015 M$_{\sun}$ to 1.21 $\pm$ 0.01 M$_{\sun}$, for a center at 1.15 $\pm$ 0.018 M$_{\sun}$. As discussed in D19, defining the optimal position for intercomparison of the Li-dips among multiple clusters can be a challenge. Clusters such as NGC 3680 and NGC 752 are inadequately populated and/or spectroscopically sampled to provide definitive boundaries for the edges of the Li-dip, though attempts have been made to approximate the location of the center of the composite cluster sample to estimate the trend with central mass with varying [Fe/H]. \citet{AT09}, assuming a symmetric Li-dip and comparing the composite cluster sample with the well-defined Hyades distribution, derived  a linear mass relation of M$_{center}$/M$_{\sun}$ = 0.4*[Fe/H] + 1.38. With [Fe/H] $= -0.04$ and a much richer sample to outline the boundaries of the Li-dip, D19 found excellent agreement between this relation and NGC 6819. If we extrapolate to [Fe/H] $= -0.55$, the center of the Li-dip in NGC 2243 should sit at 1.16 M$_{\sun}$, clearly consistent with the observations.

\begin{figure}
\figurenum{6}
\plotone{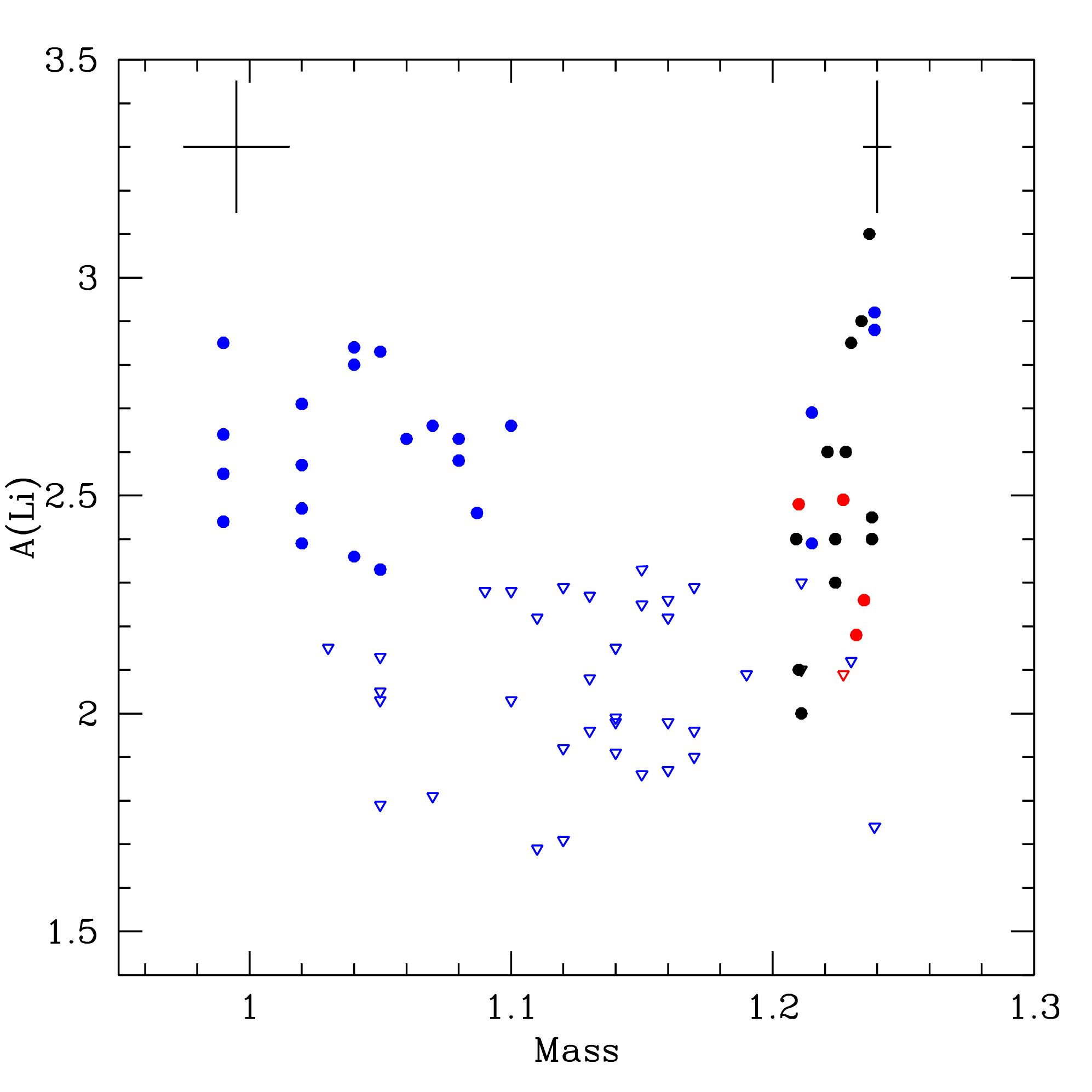}
\caption{Mass map of the stars populating the region of the Li-dip as shown in Figure 5. Symbols have the same meaning as in Figure 5. Masses are derived from a VR isochrone of 3.6 Gyr age as delineated in Figure 4. Crosses illustrate the typical error bars for A(Li) and mass at the extreme ends of the mass scale.}
\end{figure}

It should be emphasized that attempts have been made to use alternate markers to test the impact of metallicity on the evolution of the Li-dip, most notably the cool boundary of the feature (see, {\it e.g.} \citet{CU12}, FR13). The clear benefit in this approach is the survival of the lower mass boundary to a much greater age, allowing extension of the possible trend with [Fe/H] to a wider array of clusters with a greater range in metallicity, if precision spectroscopy can be obtained at the fainter luminosities of these stars. The greater age of a cluster also has an important drawback: A(Li) for stars lower in mass than the Li-dip does not remain fixed with time. A(Li) for stars at lower mass rises out of the Li-dip before plateauing and declining with decreasing mass. The rise to, and the level of, the Li plateau changes with age, making the exact definition of the boundary of the Li-dip increasingly indeterminate.

While it does have applicability to a limited range of cluster ages, with a range that depends upon [Fe/H], the A(Li) wall at the high mass boundary is usually readily identifiable if a sufficient sampling of the mass range can be accomplished. With the addition of NGC 2243, we can extend the mass boundary from the rich, well-sampled Hyades cluster at [Fe/H] = +0.15 \citep{AT09, CU17} and 1.494 M$_{\sun}$ to NGC 6819 at [Fe/H] $= -0.04$ and 1.43 M$_{\sun}$ to NGC 2243 at [Fe/H] $= -0.55$ and 1.21 M$_{\sun}$. A simple linear fit through these data generates M$_{wall}$/M$_{\sun}$ = 0.41*[Fe/H] + 1.44. We will return to this well-documented decline in mass with decreasing [Fe/H] in Section 6.

\subsection{Li Trends: Post-Main-Sequence Evolution}
Figures 7 and 8 illustrate the trends of A(Li) for stars leaving the main sequence and evolving across the subgiant branch, up the giant branch, and eventually to the red giant clump. Figure 7 separates the data by $V$ magnitude while Figure 8 details the trends with color. The turnoff region for each plot includes only stars at masses that place them outside the Li-dip. Keeping in mind the typically high S/N for the spectra at the top of the turnoff, it must be concluded that the order-of-magnitude spread in A(Li) among stars with Li detections at the top of the turnoff is real. Equally important, the range in A(Li) is not correlated with position in the CMD, either in color or luminosity. 

By the time these stars evolve to the base of the giant branch, A(Li) is further reduced to a degree that places all the stars at or below the canonical Li-rich boundary of A(Li) = 1.5 \citep{KU20}. Continuing up the giant branch, there is a general decline in A(Li) for stars with detectable Li; above the level of the red giant clump and within the clump itself, no star has detectable Li, with the upper limits set at A(Li) = 0.7 or lower.

\begin{figure}
\figurenum{7}
\plotone{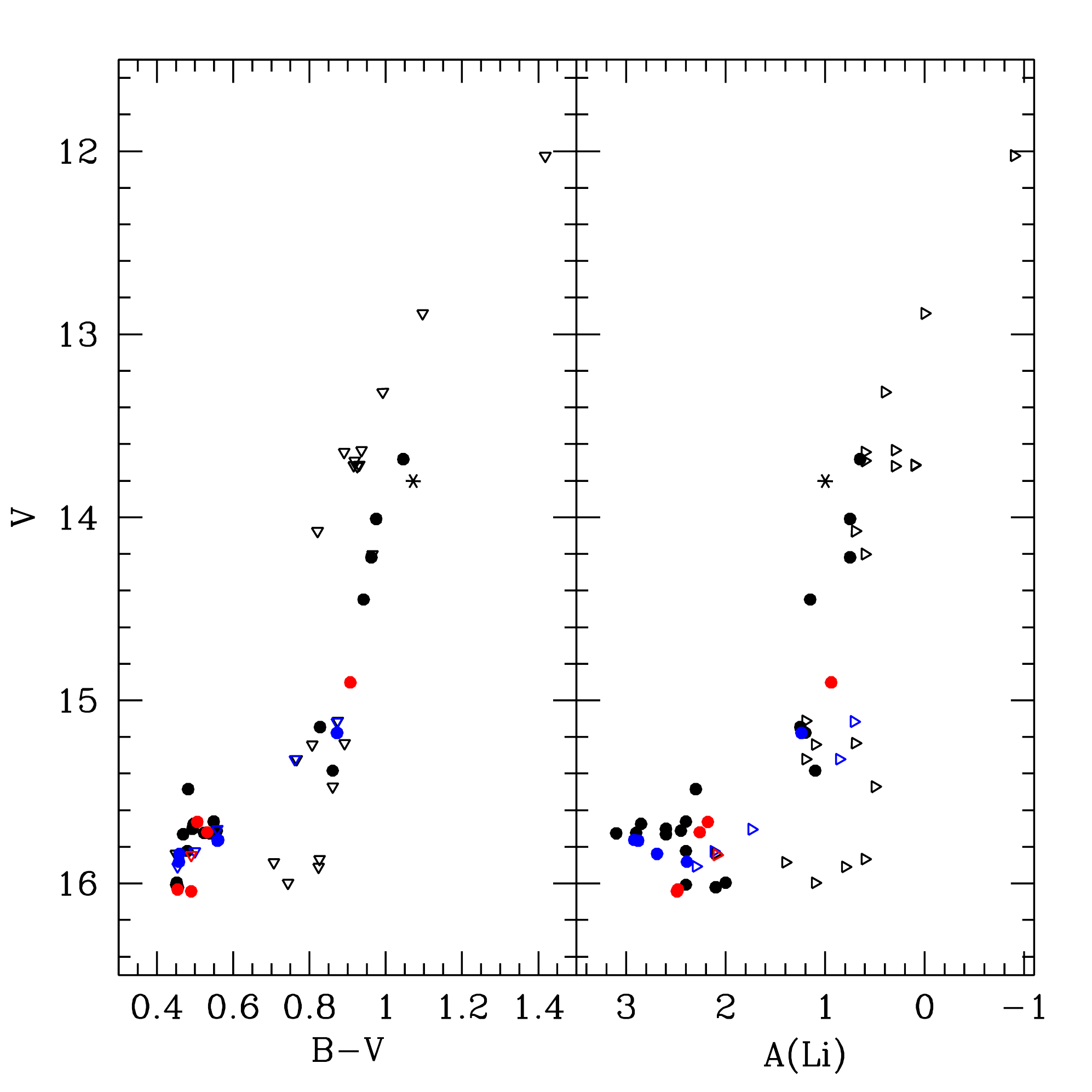}
\caption{A(Li) for stars from the top of the turnoff through the giant branch as a function of $V$. Symbols have the same meaning as in Figure 5}
\end{figure}

\begin{figure}
\figurenum{8}
\plotone{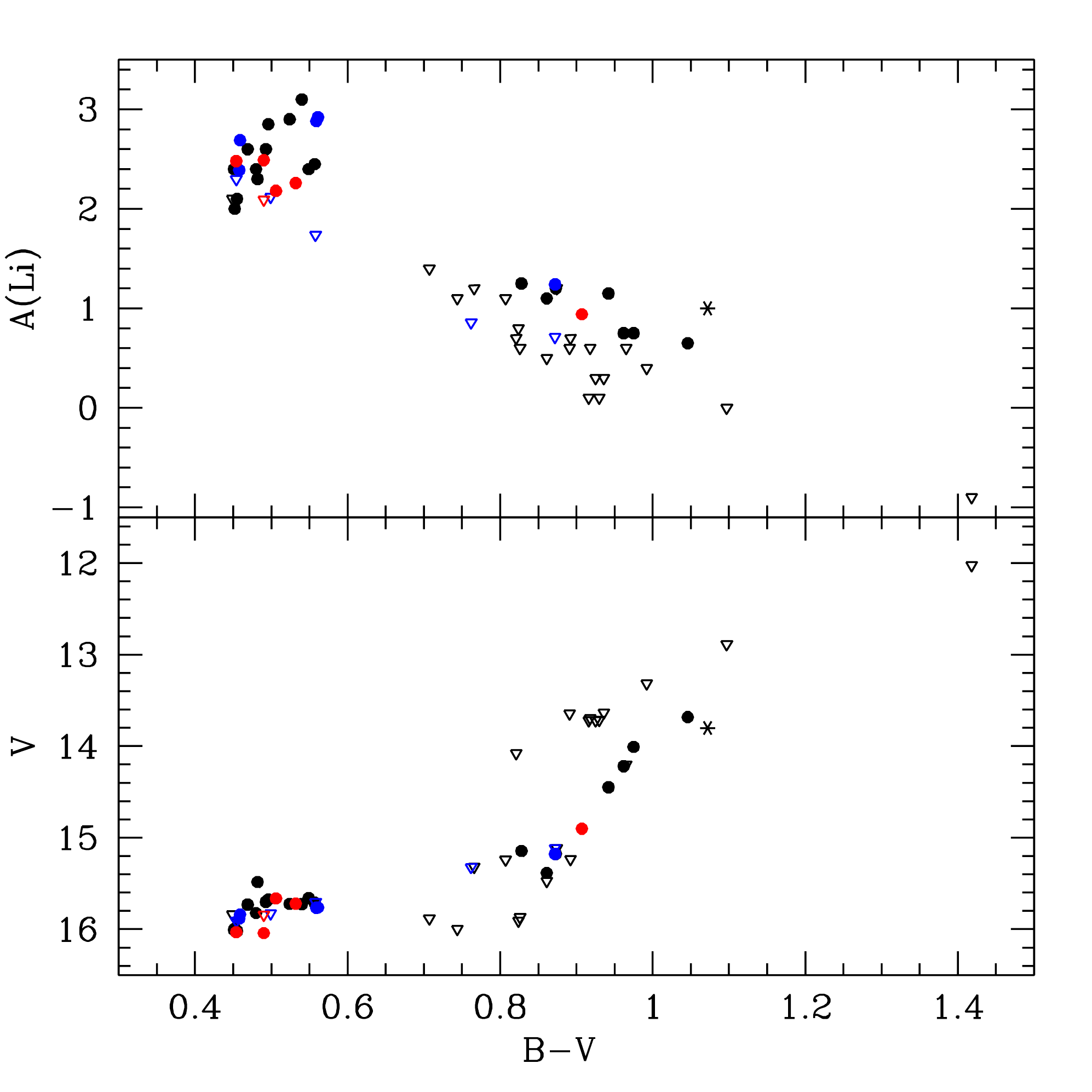}
\caption{A(Li) for stars from the top of the turnoff through the giant branch as a function of $B-V$. Symbols have the same meaning as in Figure 5, with an asterisk noting the position of W2135.}
\end{figure}

\subsection{Li Patterns: The Role of NGC 2243}

In the same way that NGC 2243 extends and strengthens the pattern of declining mass boundaries for the Li-dip with decreasing [Fe/H], its combination of low [Fe/H] and significant age place it in a unique position for testing the evolutionary trends for stars more massive than the Li-dip, while on the main sequence and beyond. Using the Hyades-Praesepe ([Fe/H] = +0.15, age = 0.65 Gyr) as a zero-point and a mix of clusters ranging in age from 1.5 Gyr (NGC 7789) to 2.25 Gyr (NGC 6819) and [Fe/H] as low as $-0.3$ (NGC 2506), D19 lay out a detailed interpretation of the manner and degree to which stars above the Li wall alter their surface A(Li) prior to leaving the main sequence and crossing the subgiant branch, where the traditionally expected impacts of a deepening convective zone and potential internal mixing dominate. In summary, it was noted that, contrary to expectation, the range in A(Li) among the stars above the wall expanded downward from the probable primordial cluster value with increasing cluster age. 

For the first test of this trend, we show in Figure 9 a comparison of the variation of A(Li) for all stars above the Li-dip from the turnoff through the giant branch for NGC 2243 (black) and NGC 2506 (red). Stars with only upper limits to the Li abundance are shown as triangles; stars with upper limits within the respective red giant clumps are three-point crosses. Being the closest in metallicity to NGC 2243 ([Fe/H] $= -0.55$), NGC 2506 ($-0.3$) should approximate the Li appearance of NGC 2243 at a younger age (1.85 Gyr vs. 3.6 Gyr), keeping in mind that the differential ``Li age'' is smaller due to the higher mass boundaries of the Li-dip for a more metal-rich cluster. Note also that the $T_{\mathrm{eff}}$ scale of the turnoff stars in NGC 2506 is hotter than that of NGC 2243 due to the younger isochronal age of the cluster.

Two striking features emerge from a comparison of the two distributions. First, among the turnoff stars in NGC 2506, 85\% have A(Li) within 0.6 dex of the maximum cluster value at the turnoff; the remaining 15\%  outside this range include both detections and upper limits. By contrast, keeping in mind the sparser sample from the turnoff of NGC 2243, there is an almost uniform distribution of stars extending from the upper bound in A(Li) to 1.1 dex  lower. For the older cluster, more than 2/3 of the turnoff stars lie below the 0.6 dex boundary in A(Li). Second, the turnoff distinction translates directly to the giant branch distribution. As discussed in \citet{AT18}, despite its moderately young age, the subgiant branch in NGC 2506 is sufficiently populated to allow delineation of the depletion of Li as stars evolve from a typical turnoff value near A(Li) $\sim$ 3.0 to a detection value of $\sim$1.4 near the base of the giant branch. By contrast, the typical value for stars leaving the main sequence in NGC 2243 is $\sim$2.5, and all stars at the base of the giant branch have only upper limits of A(Li) $\sim$ 1.2. This is clearly not the complete story since 10 stars with Li detections in NGC 2243 do overlap nicely with the giant detections within NGC 2506. Ignoring red clump stars, all of which are upper limits only in both clusters, the implication is as follows: the dominant range in A(Li) among stars leaving the main sequence in NGC 2243 is typically twice as large as that within NGC 2506 (1.1 dex vs. 0.5 dex). Depletion across the subgiant branch reduces the typical abundance at the base of the giant branch to A(Li) = 1.2 or higher for NGC 2506, leaving these stars above the detection limits of the spectroscopy. Stars that fall within the 15\% sample at the lower end of the main sequence A(Li) distribution are depleted to a comparable extent, placing them as only upper limits near the giant base, accounting for the handful of non-detections near 5300 K. For NGC 2243, clump stars aside, the ratio of non-detections to detections is 3/2 among the red giants, consistent within the uncertainties with the wide and almost uniform A(Li) distribution range among the turnoff stars. We conclude that the pattern observed among the giants merely reflects a range in extant Li in stars on the point of leaving the main sequence, convolved with one or more depletion mechanisms ({\it e.g.}, dilution and/or rotationally induced mixing) as the stars evolve along the subgiant branch and  the first-ascent giant branch.

\begin{figure}
\figurenum{9}
\includegraphics[angle=270,scale=0.30]{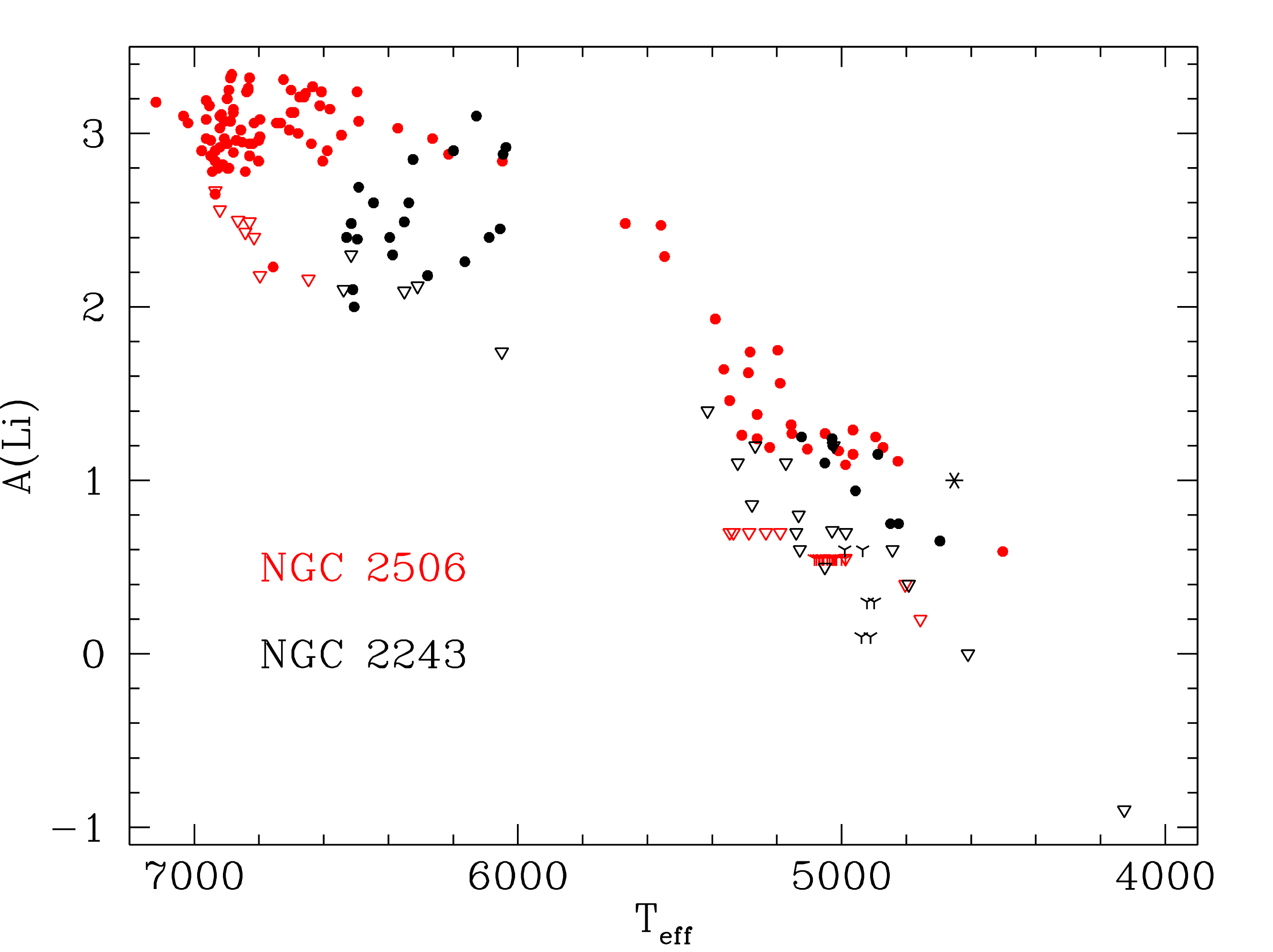}
\caption{Contrast of the A(Li) distribution with $T_{\mathrm{eff}}$ for stars more massive than the Li-dip in NGC 2506 (red) and NGC 2243 (black). Filled symbols are detections, open triangles and three-point stars are upper limits with the latter tagging clump stars. An asterisk notes the location of W2135.}
\end{figure}

To get some clarity on the potential source of the differences between NGC 2506 and NGC 2243, we turn to an older and more metal-rich cluster, NGC 6819.
While only 0.4 Gyr older than NGC 2506, from the standpoint of ``Li age'', the cluster is significantly more advanced due to its higher metallicity ([Fe/H] $= -0.04$). This positions the Li-dip within a mass range 0.1 M$_{\sun}$ higher than in NGC 2506; its turnoff stars (above the Li-dip) are in a closer evolutionary state relative to the Li-dip to those in NGC 2243 rather than NGC 2506. As a simplistic means of quantifying this evolutionary phase, we note that the range in magnitude for stars above the Li-dip in the three clusters of interest is, binaries aside, 0.6 mag, 1.0 mag, and $>1.4$ mag for NGC 2243, NGC 6819, and NGC 2506, respectively. The lower bound is given for NGC 2506 because the exact location of the wall of the Li-dip in this cluster has not yet been determined \citep{AT18}. Figure 10 reveals the NGC 6819 analog to Figure 9, with the notable change that no distinction is made between upper limits in A(Li) for clump giants relative to first-ascent giants since so few giants of any category in NGC 6819 exhibit Li detections.

\begin{figure}
\figurenum{10}
\includegraphics[angle=270,scale=0.30]{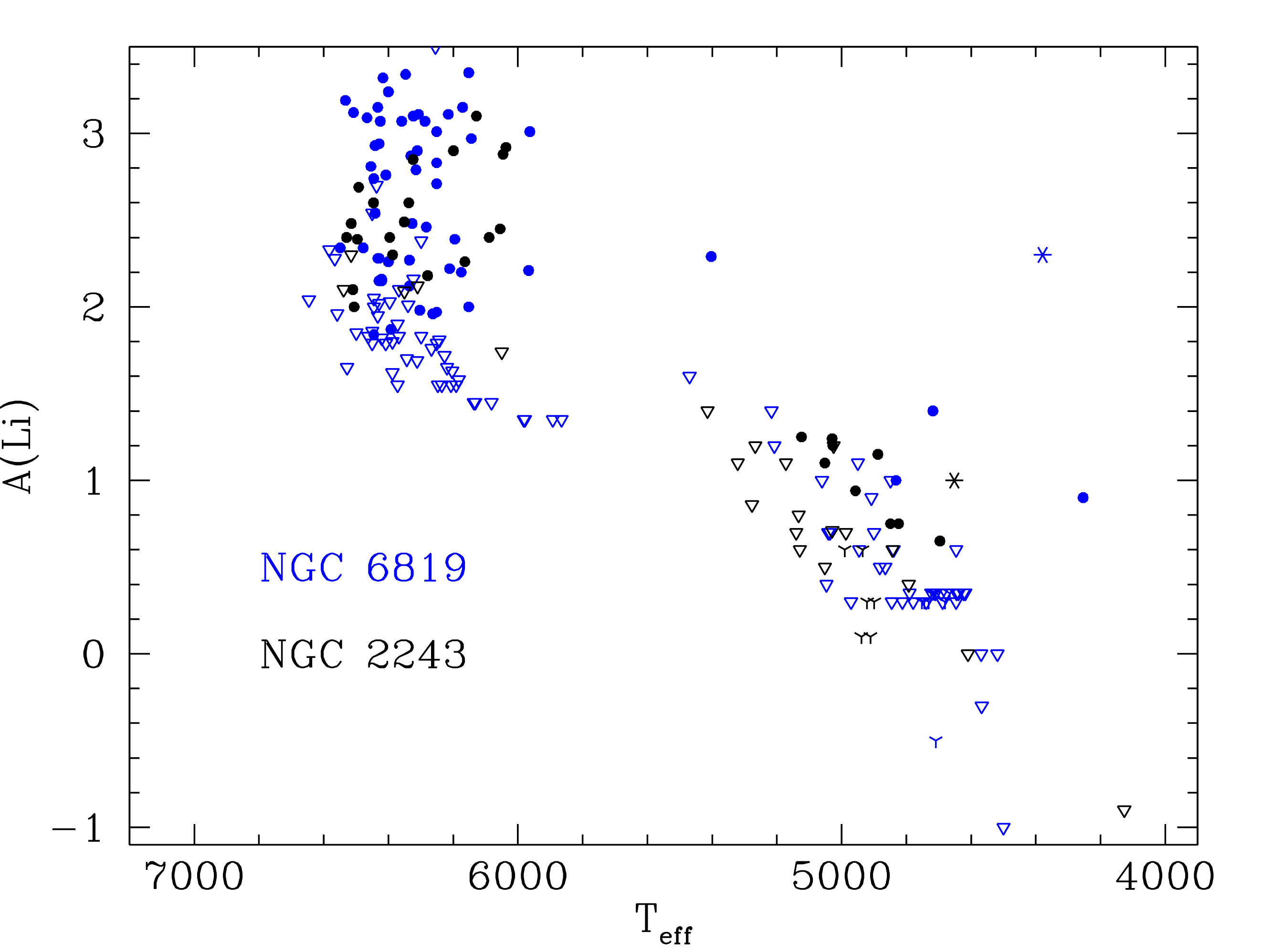}
\caption{Same as Figure 9 for NGC 6819 (blue) and NGC 2243 (black). Clump stars in NGC 6819 are, like the majority of giants, all upper limits. No distinction is made between first-ascent and clump giants in NGC 6819. Asterisks note the presence of Li-rich giants W2135 in NGC 2243 and W7017 in NGC 6819.}
\end{figure} 

Comparison of Figure 10 with Figure 9 demonstrates that the NGC 6819 turnoff distribution bears a much stronger similarity to NGC 2243 than NGC 2506. Keeping in mind the statistically richer sample of NGC 6819, the range in A(Li) among detections is 1.4 dex; including upper limits 
the range extends to 2.0 dex. 
More relevant, the distribution is even more heavily weighted toward lower A(Li) than in NGC 2243; 30\% of the sample lies within 0.6 dex of the cluster limit but 63\% are more than 1 dex below the cluster limit and are dominated by upper limits rather than detections (D19). 
Because the range in A(Li) among the stars hotter than 5800 K is uncorrelated with either luminosity or $T_{\mathrm{eff}}$ above the Li-dip, stars reaching the base of the giant branch should exhibit a comparable spread in A(Li), assuming the depletion mechanism on the subgiant branch applies equally to all stars independent of their initial A(Li) upon entering the subgiant branch. The ubiquity of upper limits among the giants in NGC 6819 is consistent with this prediction given the turnoff distribution weighted toward depleted Li. Clearly some stars retain enough Li to fall within the detection limits, supposedly the descendents of the stars leaving the main sequence with Li relatively unchanged from the primordial cluster value. Only 5 stars cooler than 5600 K in NGC 6819 have detectable Li. Of the two stars with A(Li) greater than 2.0, one lies on the subgiant branch between the turnoff and the base of the giant branch while the second is WOCS7017, a Li-rich giant now known to be a clump star of anomalously low mass \citep{AT13, CA15, HA17}. 
 
Given Figs. 9 and 10, the question remains: which stellar parameter(s) control the turnoff distribution of A(Li) as a function of age and metallicity and thus the trend among the giants? Following the illuminating discussion of D19, we show in Figure 11 the range of $v_{rot}$ as a function of the star's position above the wall of the Li-dip, a more restricted version of their Figure 11, including only the three clusters under discussion, NGC 2243, NGC 6819, and NGC 2506 and plotting only stars outside the Li-dip. $\Delta V$ is the distance in magnitudes a star lies positioned above the wall of the Li-dip, in the sense ($V_{star} - V_{wall}$). The velocity range within the youngest cluster (from a ``Li age'' standpoint) is just under 100 
km-sec$^{-1}$; for both NGC 6819 and NGC 2243, the range is virtually identical, with only one star in each cluster above 25 km-sec$^{-1}$.
 
When placed in the context of all the clusters detailed in D19, it is concluded that the primary control of the Li distribution with age for stars above the Li wall is the rotational spindown of stars from an initial range close to or greater than 100 km-sec$^{-1}$ (NGC 7789 and NGC 2506) to less than 25 km-sec$^{-1}$ (NGC 6819 and NGC 2243). The spindown mechanism, while still on the main sequence, is either correlated with and/or triggers the variable decline in Li among stars with a significant range in primordial $v_{rot}$, generating a smaller range in $v_{rot}$ with increasing age, but an increasing range downward in A(Li) from the primordial cluster value. 

It should be emphasized that the A(Li) distribution for stars above the wall prior to rotational spindown on the main sequence need not be a single-valued peak set at the initial cluster value. Diffusion of Li, with the degree of depletion defined by the level of rotational mixing, could produce a range of A(Li) among turnoff stars prior to rotational spindown, as seen, for example, among the turnoff stars in NGC 2506 in Figure 9. The spindown would then initially raise the atmospheric Li content as Li from stable layers below the surface is mixed to the surface, supposedly returning the star to its primordial value before enhanced rotational mixing drives the retrieved Li content to even greater depth and destruction. This process could be indicated by the Li trend among the subgiants in Figure 8 between $B-V$ = 0.45 and 0.6, but the sample remains too small and the data sources too heterogeneous to reach any definitive conclusion.

We have shown that both
age and metallicity are important parameters for Li depletion.  Two
clusters of roughly similar age (NGC 2506 and NGC 6819) but different
metallicity show substantially different subgiant and giant Li patterns,
with NGC 2506 being the more orderly of the two.  But it's also true that
among the two metal-poor clusters (NGC 2506 and NGC 2243), the older one
(NGC 2243) is less orderly, and is more similar to NGC 6819.  The older,
metal-poor cluster bears more resemblance to the younger, metal-rich
cluster than it does to the younger, metal-poor cluster.
One straightforward implication of this trend is that, assuming that the degree of post-main-sequence Li depletion is independent of the initial Li abundance upon entering the subgiant branch, the range in A(Li) among giants at a given phase of evolution will depend strongly upon the mass and metallicity of the stars leaving the main sequence, {\it i.e.} classification of a giant as Li-rich or Li-poor requires boundaries that vary with mass, metallicity, and age \citep{TW20}. 

\begin{figure}
\figurenum{11}
\plotone{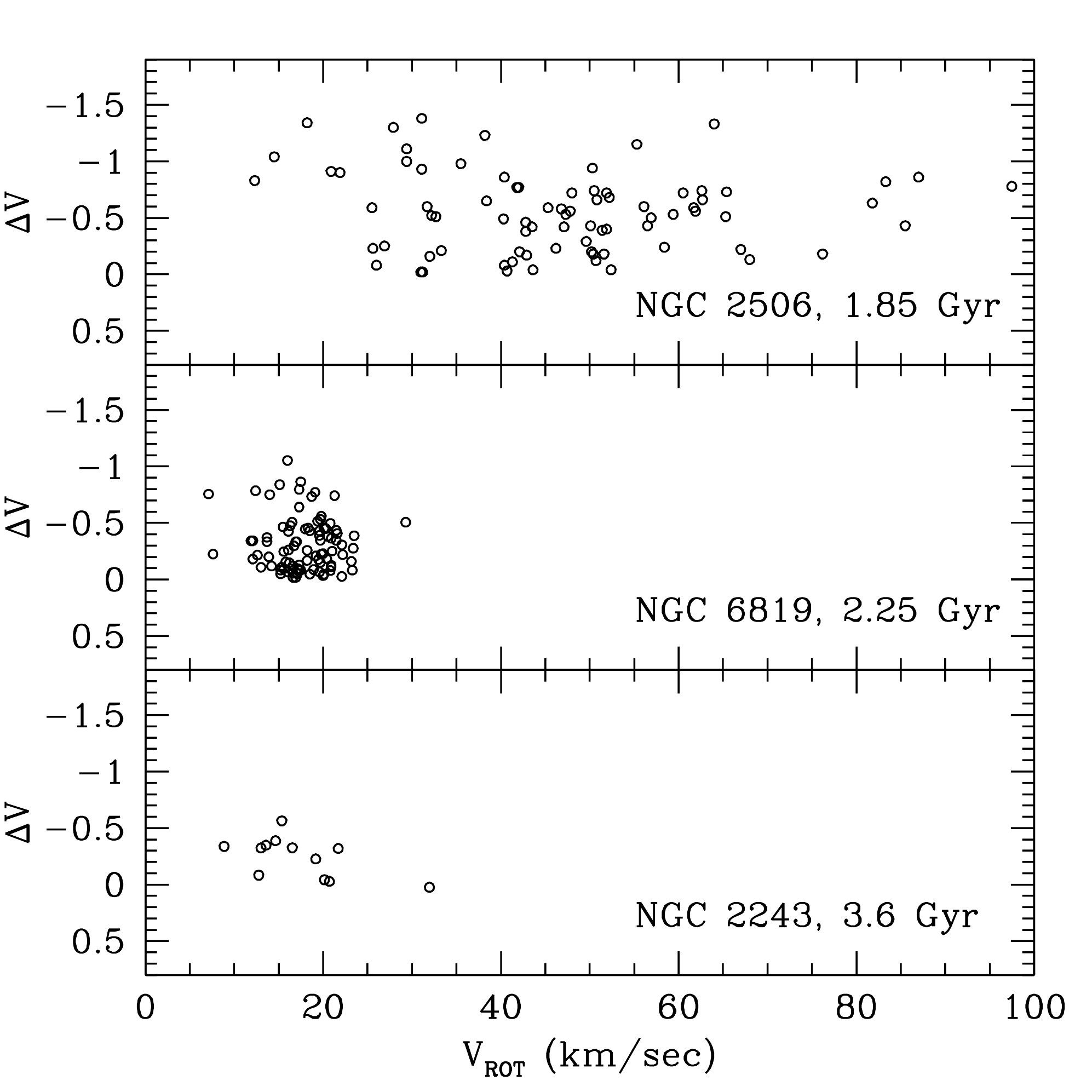}
\caption{Distribution in $v_{rot}$ for stars above Li-dip for NGC 2506, NGC 6819, and NGC 2243. From the standpoint of Li evolution, NGC 6819 and NGC 2243 are of comparable age and much older than NGC 2506, in contrast with the listed isochronal ages. $\Delta V$ = 0 is defined by the wall of the Li-dip.}
\end{figure}

\section{NGC 2243 Summary and Galactic Implications}
High dispersion spectra have been obtained of 42 stars within the field of the metal-deficient open cluster, NGC 2243. From both astrometry and radial velocities, 32 cluster members have been identified and their spectra analyzed. From 29 likely single stars, the overall cluster [Fe/H] is well-defined at $-0.54 \pm$ 0.11, where the error describes the median absolute deviation about the sample median value, leading to internal precision for the cluster [Fe/H] below 0.03 dex. A much more restricted sample of lines for $\alpha$-elements produces effectively scaled-solar abundances, though the uncertainties are larger. With the metallicity and  previously determined reddening in hand, the cluster age and distance are obtained by comparison to an appropriate set of isochrones, generating an age and apparent distance modulus of 3.6 $\pm$ 0.2 Gyr and $(m-M)$ = 13.2 $\pm$ 0.1, respectively. The former value places the cluster in an age range comparable to that of M67 while the latter parameter is in excellent agreement with the predicted parallax-based calculation from Gaia DR2, after correction for the probable zero-point parallax error for a cluster at the large distance of NGC 2243.

Since the primary goal of the investigation is the delineation of the Li abundance among stars within the cluster from the main sequence through the giant branch, as well as placing the cluster within the context of Li evolution among the open cluster population with age and metallicity, the Hydra sample was expanded through the inclusion and analysis of high S/N, high resolution cluster spectra from the VLT archive, and the addition of Li abundances for members only available from the published data of \citet{FR13, HI00}, boosting the total cluster sample to over 100 stars. The expanded sample has the greatest impact in defining the variation of Li among stars more massive than the Li-dip, from the top of the turnoff through the giant branch. The Li-dip remains clearly defined; 
what is accomplished by the doubling of the turnoff stars more massive than the dip, all but one of which are detections rather than upper limits, is the clear indication that the cluster upper limit to A(Li) is at least 2.9, with one star positioned near 3.1.
This is quite similar to the peak value of 2.96 measured by \citet{ra20} despite a higher overall [Fe/H] estimate of $-0.38$ from Gaia ESO pipeline analysis; references within \citet{ra20}. \citet{ra20} also correctly note upper turnoff stars ``may have already undergone some post-MS dilution'' to their surface Li abundances making a detection of {\it the} peak A(Li) value stochastically vulnerable to sample size and selection.

Moreover, the spread in A(Li) among stars on the verge of entering the subgiant branch is at least 1.1 dex and could be greater than 1.3 dex. This spread is crucial because it is not dependent upon specific position within the turnoff region, {\it i.e.} stars occupying virtually identical locations within the CMD differ in their A(Li) by 0.9 dex. This spread translates directly into a large spread among giants ascending the giant branch for the first time. Excluding the post-He-flash members of the red clump, all of which have only upper limits to A(Li), the majority of the red giants exhibit only upper limits below A(Li) = 1.5, the nominal value predicted for stars of solar Li upon arrival at the base of the giant branch. (Note, the limits become more restrictive with increasing luminosity due to the declining $T_{\mathrm{eff}}$ among brighter giants and should not be interpreted as evidence for declining A(Li) with evolution up the giant branch; only the Li detections can illuminate this pattern.) This trend is consistent with the pattern found in the more metal-rich and numerically richer but younger cluster, NGC 6819. The implication is that, due to the shift in the Li-dip to lower mass at lower [Fe/H], the evolutionary phase of a cluster with respect to Li evolutionary pattern, the ``Li age'', requires a more metal-deficient cluster to attain a greater age to approach the same ``Li age'' as a metal-rich cluster.

Within the discussion of Li evolution among stars, the role of the main sequence/turnoff stars above the wall has been twofold. First, by evaluating the sensitivity of the wall to changes in mass, metallicity, and age, some insight could be gained into the physical process(es) triggering the depletion of Li among main sequence stars both within the Li-dip and beyond. As discussed in detail in D19 and confirmed with the addition of NGC 2243, a controlling factor in the evolutionary history of Li for any star, whether above the wall, within, or below the Li-dip appears to be the spindown mechanism of the star. The key, but not necessarily the only, difference among the stars at varying masses on the main sequence is whether the decline in rotation occurs rapidly, {\it e.g.} during the pre-main-sequence phase, or more gradually (and/or nonlinearly) during main-sequence evolution. The challenge in identifying such trends comes from the fact that the degree of Li depletion is likely to be correlated with the degree of rotational decline, but what one measures is an instantaneous snapshot of the current rotational speed with no {\it a priori} knowledge of the star's initial rate upon arrival at the main sequence, though statistical distributions of the kind shown in Figure 11 (and in Figure 11 of D19) can help constrain the likely degree of change. If correct, the spindown pattern among the cluster samples and the long-standing decline in the mass location of the Li-dip with decreasing [Fe/H] points toward a metallicity dependence in the rotational evolution of the main sequence stars, through an initial $v_{rot}$ distribution with mass which depends upon [Fe/H] and/or a spindown mechanism whose effectiveness changes with [Fe/H], a question whose resolution, while intriguing, is well beyond the scope of the current sample.

Second, due to the rapid establishment of the Li-dip within clusters by the age of the Hyades/Praesepe \citep{CU17} and the long-term decline in Li for stars lower in mass than the Li-dip (see, {\it e.g.} \citet{CU12, BO18, AT18} among many others), the stars above the wall have represented a potential pristine indicator of the primordial cluster Li abundance, a clear reference marker against which changes to lower mass stars could be calibrated, again assuming that the higher mass stars were minimally affected by processes such as diffusion. It is now apparent that as a cluster ages the Li abundance of the stars above the wall does evolve downward, with the degree of depletion growing with age due to the declining range in $v_{rot}$, usually well beyond the range defined by the statistical uncertainty in the determination of individual A(Li). Not only does this weaken the value of these stars in defining the original cluster abundance, among the stars above the wall it essentially recreates the Li-dip, though at a slower pace and less effectively (D19). The impact on the giant branch is crucial. The masses of the stars reaching the base of the giant branch with severely reduced A(Li) are no longer defined by the mass of the Li wall, but now extend significantly higher. For the numerically rich cluster NGC 6819 at [Fe/H] $= -0.04$, Li-depletion is well underway among stars leaving the main sequence with minimal masses of 1.57 M$_{\sun}$, even though the Li wall doesn't occur until one reaches 1.43 M$_{\sun}$. Clearly these boundaries will shift downward will decreasing metallicity. Because the level of the Li plateau among lower mass stars beyond the cool edge of the Li-dip on the main sequence will decrease with time \citep{CU12}, the majority of the red giants older than $\sim$2 Gyr will be well below the Li-rich boundary of A(Li) = 1.5, long before they reach the red giant tip and the eventual red giant clump. (Again, the age boundary will depend strongly upon [Fe/H]. For example, in NGC 2243, Li-depleted stars above the wall are already leaving the main sequence, despite the fact that the stars defining the Li wall won't begin to populate the subgiant branch until the cluster is 4.1 Gyr old.) Thus, not only is it likely that any lower mass red clump star with a detectable A(Li) should be classified as Li-rich \citep{KU20}, a population of lower mass stars on the first-ascent giant branch now tagged as Li-poor may actually show evidence of post-main-sequence Li enhancement prior to arrival at the red giant tip \citep{TW20,SU20}.

We close the investigation of NGC 2243 with a discussion of the broader impact of the delineation of the Li-dip with metallicity on our understanding of Galactic Li evolution. The evolution of stars above the wall, {\it i.e.} on the warm side of the Li-dip, and the metallicity-dependent change in the mass boundaries of the Li-dip have subtle but relevant impacts on the interpretation of the Li distribution among field stars. In a recent analysis of the exquisite GALAH database for F and G dwarfs, \citet{GA20} identify and analyze over 100000 stars between [Fe/H] $= -3$ and +0.5 and $T_{\mathrm{eff}}$ between 5900 K and 7000 K. After sorting the stars into three Li categories, ``warm" (above the wall), within the Li-dip, and ``cool" (below the Li-dip), the trends with [Fe/H] for the ``warm" and ``cool" stars were compared. As expected, for stars with [Fe/H] below -1.0, only ``cool" stars were present due to the evolution of ``warm" stars off the main sequence by the age typical of these metal-poor dwarfs. The approximately uniform A(Li) for these stars is consistent with Li abundances for stars occupying the Spite plateau \citep{SP82a, SP82b}. Between [Fe/H] $= -1.0$ and $-0.5$, a ``warm" sample of 117 stars reappeared and scattered about a mean value of A(Li) = 2.69 $\pm$  0.06, statistically identical with the predicted Big Bang Nucleosynthesis value of 2.75. With increasing [Fe/H] above $-0.5$, A(Li) among both the ``warm" and the ``cool" stars exhibits a steady increase, implying a contribution from Galactic nucleosynthesis superposed upon the initial undiluted primordial Li abundance defined by the stars at [Fe/H] $= -0.5$ to $-1.0$. 

However, the above interpretations of the ``warm" star patterns above [Fe/H] $= -1.0$ are straightforward only if the assumption that these stars retain their initial Li abundance prior to entering the subgiant branch is appropriate. If the trend among the stars above the wall (``warm") with increasing age as defined by the clusters discussed in D19 and extended to [Fe/H] $= -0.5$ by NGC 2243 is correct, the mean of the A(Li) distribution of the ``warm" stars increasingly underestimates any initial cluster value with increasing age due to Li depletion triggered by rotational spindown. Thus, taking the 20 stars above the wall in NGC 2243 as representative of a sample of field ``warm" stars at [Fe/H] $= -0.5$, the mean A(Li) would be 2.52 $\pm$ 0.06 (sem), even though the cluster limiting value appears to lie between A(Li) = 2.85 and 3.1. One could minimize this effect if the sample selected is dominated by younger stars of the same mass, but the probability of finding such stars as [Fe/H] declines from $-0.5$ to $-1.0$ seems unlikely.

Equally important is the effect of [Fe/H] on the Li-dip boundary, {\it i.e.} the defining lines for ``warm" and ``cool" stars. Fig. 12 illustrates the competing effects of basic stellar evolution with the potential development of the boundaries of the Li-dip. Using the VR isochrones defined with [$\alpha$/Fe] set at +0.4, the mass range of the stars populating the CMD turnoff at an age of 12.5 Gyr with decreasing [Fe/H] is shown as a pair of solid lines. The mass range of the turnoff is defined from the bluest point at the of the main sequence CMD to a position 0.10 mag redder in $B-V$, the latter boundary selected to avoid stars normally classified as subgiants. The masses are set using isochrones with an age of 12.5 Gyr across the board. Clearly, shifting the models to younger ages would slide the relations to a systematically higher mass. By contrast, the dashed lines show the mass boundaries of the Li-dip, created by simple extrapolation of the linear trends defined by multiple clusters from [Fe/H] $= -0.5$ (NGC 2243) to +0.45 (NGC 6253) \citep{CU12}. We note that the lower mass boundary of the Li-dip as applied in Fig. 12 is not defined by the mass of the cool star plateau \citep{CU12, FR13} beyond the Li-dip, but the mass where A(Li) initially rises to a level of consistent detectability. In the case of NGC 2243, this shifts the lower mass boundary from 1.03 M$_{\sun}$ \citep{FR13} to 1.09 M$_{\sun}$ (Figure 6).

As expected, the simplistic extrapolation of the Li-dip mass range crosses the globular cluster mass trend between [Fe/H] $= -1.25$ and $-1.50$, in contradiction with the absence of the Li-dip within globular cluster CMDs and the existence of the Spite plateau. An obvious solution to the contradiction is that the mass/metallicity slope flattens below [Fe/H] $= -0.5$, preventing the Li-dip from ever crossing the evolutionary trend with lower metallicities except among younger isochrones. Because the mass/metallicity relation is based upon clusters with scaled-solar abundances, it is possible that it is inapplicable to stars/clusters with enhanced [$\alpha$/Fe], {\it i.e.} metallicity should refer to [m/H] rather than [Fe/H]. If the typical star below [Fe/H] $= -0.5$ has [$\alpha$/Fe] = +0.4, as assumed for the isochrones, the mass boundaries should be boosted by as much as 0.14 to 0.16 M$_{\sun}$ at the ``cool' and ``warm" boundaries, pushing the crossover of the two trends to [Fe/H] below $-2.5$. 

The relevance of these qualitative trends is illustrated by the sequence of ages (Gyr) plotted as a function of [Fe/H] in Figure 12. The blue terms define the last age at which a star of the given [Fe/H] can be found on the ``warm" side of the Li-dip, assuming the mass boundary is defined by extrapolation of the linear trend above [Fe/H] $= -0.5$, not including the possible role of [$\alpha$/Fe] = +0.4 within the metallicity. The red points define the same age marker, but include the enhanced $\alpha$ elements in the mass boundary determination, {\it i.e.} M/M${\sun } = -0.4$*[m/H] + constant, for the ``warm" stars. Two points are immediately obvious. In either case, there is a significant age bias in the selection of observable stars. The more metal-rich the star, the younger it has to be to be included in a ``warm" star sample. Second, mechanisms used to alter the slope of the mass/metallicity relation by making it shallower or by shifting it to higher mass limits with metallicity, as done via alteration of the definition of metallicity in Figure 12, invariably move the age boundary for the ``warm" stars to a younger age, creating an even more extreme selection bias within any ``warm" star sample from the field. 

\begin{figure}
\figurenum{12}
\plotone{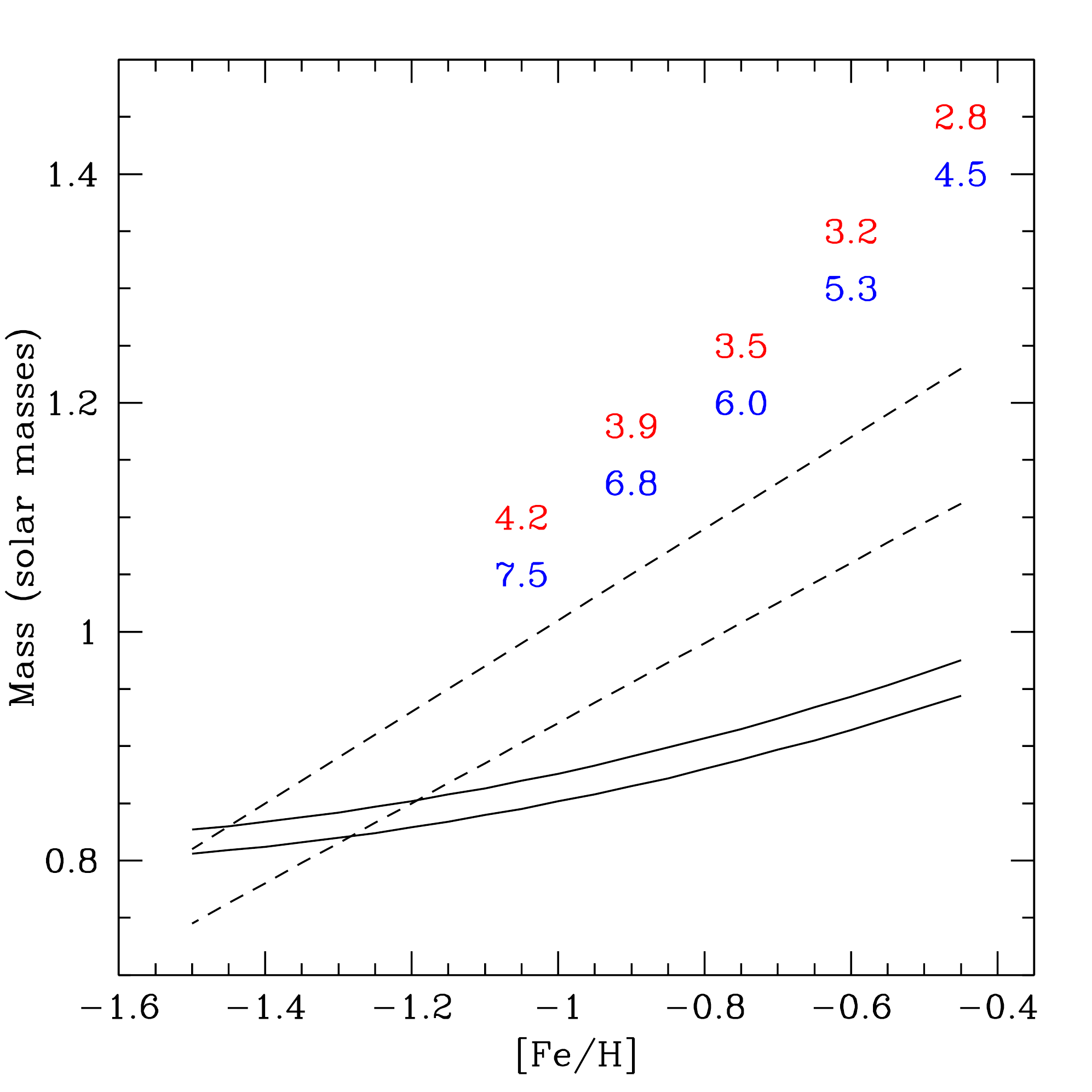}
\caption{The mass range of the Li-dip (dashed lines) compared to the mass range for the turnoff stars in a cluster of age 12.5 Gyr (solid lines) as a function of [Fe/H]. Isochrones are assumed to be [$\alpha$/Fe] = 0.4. Age (Gyr) at which the Li wall evolves to the subgiant branch is given in blue for metallicity defined as [Fe/H], red for [m/H], $\alpha$ enhancement included.}
\end{figure}

The selection bias due to age limitations for populating the ``warm" side of the Li-dip, coupled with Li depletion among ``warm" stars prior to entering the subgiant branch, has already been seen among stars at the metal-rich end of the scale. A number of studies have claimed that the limit to A(Li) among stars of supersolar metallicity declines as [Fe/H] increases (see, e.g., \citet{fu18,gu19,ben20,stou20}). 
This apparent trend is readily understood if the Li-dip boundaries plotted in Figure 12 are extended to [Fe/H] = +0.5, using isochrones with scaled-solar abundances. From our mass/[Fe/H] relation, at [Fe/H] $= -0.04$ and +0.43 the mass boundaries for the wall are located at 1.42 and 1.61 M$_{\sun}$, respectively. Using the same definition as earlier, these mass boundaries enter the subgiant branch at an age of 3.1 and 2.4 Gyr, respectively. The former metallicity is representative of NGC 6819 (age = 2.3 $\pm$ 0.1 Gyr (D19)) with a spread of A(Li) among its warm stars of 1.4 dex. The latter metallicity is representative of NGC 6253 \citep{AT10, CU12} (age = 3.0 $\pm$ 0.4 Gyr). Stars leaving the turnoff in NGC 6253 are clearly emerging from within the Li-dip; of the 12 stars on the warmer side of the Li-dip and across the subgiant branch, only 2 have detections and both of these are below A(Li) = 2.1, even though the stars populating the cool star plateau beyond the Li-dip almost reach A(Li) = 2.8. 

If the spreads in A(Li) among the ``warm" stars in NGC 2243 and NGC 6819 seen prior to the evolution of their walls onto their subgiant branch are typical of clusters of higher metallicity of comparable ``Li-age", one would expect to see a significant range in A(Li) among ``warm" stars at [Fe/H] near +0.4 between the ages of approximately 1.6 and 2.3 Gyr. Thus, for supersolar metallicity the initial cluster A(Li) would only remain detectable as a mean value among the ``warm" stars at ages below $\sim$ 1.6 Gyr, with the boundary rising as [Fe/H] declines to solar. For ages between this boundary and the age when the wall enters the subgiant branch, the initial A(Li) may still be detectable, but only as an upper limit to the sample. This pattern is totally consistent with the Li analysis of 18 clusters, 7 of which have [Fe/H] $> 0$,  by \citet{ra20}. With one cluster at 1.4 Gyr and all others at 1.0 Gyr or less, \citet{ra20} find no decrease in A(Li) with increasing [Fe/H]. In fact, the clusters with higher metallicity have the highest A(Li). 

\acknowledgments
NSF support for this project was provided to BJAT and BAT through NSF grant AST-1211621, and to CPD through NSF grants 
AST-1211699 and AST-1909456. The assistance of KU undergraduate Steven Smith was instrumental in the early stages of this project; Dr. Donald Lee-Brown assisted with observations and development of the ANNA code. 

This research has made use of the WEBDA database, operated at the Department of Theoretical Physics and Astrophysics of the Masaryk University. 
This research has made use of the services of the ESO Science Archive Facility 
and is based on observations collected at the European Southern Observatory under the public {\it Gaia} ESO Survey Program.

\facility{WIYN: 3.5m}

\software{IRAF \citet{TODY}, ANNA \citet{LB17, LB18}, MOOG \citet{SN73}, LACOSMIC \citet{VD01} }

\end{document}